\documentclass[twocolumn,runningheads]{svjour2}
\smartqed  % flush right qed marks, e.g. at end of proof
\usepackage{graphicx}
\usepackage{mathptmx}   % use Times fonts if available on your TeX system
\usepackage[authoryear]{natbib}
\def\note #1]{{\bf #1]}}
\newcommand{\dd}{{\rm d}}
\newcommand{\ADIPLS}{{\small ADIPLS}}
\newcommand{\FILOU}{{\small FILOU}}
\newcommand{\GRACO}{{\small GraCo}}
\newcommand{\LOSC}{{\small LOSC}}
\newcommand{\NOSC}{{\small NOSC}}
\newcommand{\OSCROX}{{\small OSCROX}}
\newcommand{\POSC}{{\small POSC}}
\newcommand{\ROMOSC}{{\small LNAWENR}}
\newcommand{\PULSE}{{\small PULSE}}

\newcommand{\ASTEC}{{\small ASTEC}}
\newcommand{\CESAM}{{\small CESAM}}

\journalname{Astrophysics and Space Science (CoRoT/ESTA Volume)}
\begin{document}

\title{Inter-comparison of the g-, f- and p-modes calculated using
different oscillation codes for a given stellar model %\thanks{Grantsor other notes
}


\author{A.~Moya \and J.~Christensen-Dalsgaard \and S.~Charpinet \and
  Y.~Lebreton \and A.~Miglio \and J.~Montalb\'an \and M.J.P.F.G.~Monteiro \and
  J.~Provost \and I.W.~Roxburgh \and R.~Scuflaire \and J.C.~Su\'arez \and
  M.~Suran
%etc.
}


\institute{A. Moya \and J.C. Su\'arez \at
			Instituto de Astrof\'{\i}sica de Andaluc\'{\i}a- CSIC \at
            Cno. Bajo de Huetor, 50, Granada, Spain\\
            \email{moya@iaa.es}
    \and 
    	    J. Christensen-Dalsgaard \at
	    Institut for Fysik og Astronomi, og Dansk AsteroSeismisk Center,
	    Aarhus Universitet, Denmark
    \and
        S. Charpinet \at
        Observatoire Midi-Pyr\'en\'ees, France
    \and
        Y.~Lebreton \at
        Observatoire de Paris, GEPI, CNRS UMR 8111,
        Meudon, France
    \and A. Miglio \and J.~Montalb\'an \and R. Scuflaire \at
        Institut d'Astrophysique et Geophysique,
	    Universit\'e de Li\`ege, Belgium
    \and M.J.P.F.G.~Monteiro \at
        Centro de Astrof\'{\i}sica da Universidade do Porto and 
        Departamento de Matem\'atica Aplicada da Faculdade de Ci\^encias,
        Universidade do Porto, Portugal
    \and
        J. Provost \at
        Observatoire de la Cote d'Azur, Nice, France
    \and
        I. Roxburgh \at
        Astronomy Unit, Queen Mary, University of London, UK
    \and
        M.~Suran \at
  	    Astronomical Institute of the Romanian Academy, Romania
}

\date{Received: date / Accepted: date}

\maketitle
%*******************************************************************
\begin{abstract}

In order to make asteroseismology a powerful tool to explore stellar
interiors, different numerical codes should give the same oscillation
frequencies for the same input phy\-sics. Any differences
found when comparing the numerical values of the eigenfrequencies will
be an important piece of information regarding the numerical structure
of the code.  The ESTA group was created to analyze the non-physical
sources of these differences. The work presented in this report is a
part of Task~2 of the ESTA group. Basically the work is devoted to
test, compare and, if needed, optimize the seismic codes used to
calculate the eigenfrequencies to be finally compared with
observations. The first step in this comparison is presented here. The
oscillation codes of nine research groups in the field have been used
in this study. The same physics has been imposed for all the codes in
order to isolate the non-physical dependence of any possible
difference. Two equilibrium models with different grids, 2172 and 4042
mesh points, have been used, and the latter model
includes an explicit modelling of semiconvection just outside
the convective core. 
Comparing the results for these two models illustrates the effect of the
number of mesh points and their distribution in particularly critical
parts of the model, such as the steep composition gradient outside the
convective core.
%These two models also present the important
%difference of a semi-convection treatment close to the boundary of the
%convective core.
A comprehensive study of the frequency differences
found for the different codes is given as well. These differences are
mainly due to the use of different numerical integration schemes. The
number of mesh points and their distribution are crucial for
interpreting the results. The use of a second-order integration scheme
plus a Richardson extrapolation provides similar results to a
fourth-order integration scheme. The proper numerical description of
the Brunt-V\"ais\"al\"a frequency in the equilibrium model is also
critical for some modes. This influence depends on the set of the
eigenfunctions used for the solution of the differential equations. 
An unexpected result of this study is the high sensitivity of the
frequency differences to the inconsistent use of values of the gravitational
constant ($G$) in the oscillation codes,
within the range of the experimentally determined ones,
which differ from the value used to compute the equilibrium model.
This effect can provide differences for a
given equilibrium model substantially larger than those resulting from
the use of different codes or numerical techniques; the actual
differences between the values of $G$ used by the different codes account for
much of the frequency differences found here.

%\keywords{First keyword \and Second
%keyword \and More} \PACS{First \and Second \and More}
\end{abstract}

%-------------------------------------------------------------------
\section{Introduction}\label{intro}

Asteroseismology is at present being developed as an efficient
instrument in the study of stellar interiors and
evolution. Pulsational frequencies are the most important
asteroseismic observational inputs.  It is evident that a meaningful
analysis of the observation, in terms of the basic physics of stellar
interiors which is the ultimate target of the investigation, requires
reliable computation of oscillation frequencies for specified physics.
This is a two-step process, involving first the computation of stellar
evolutionary models and secondly the computation of frequencies for
the resulting models.  Lebreton et al. (this volume) provide an
overview of the tests of stellar model calculations.  Here we consider
the computation of the oscillation frequencies.

An evident goal is that the computed frequencies, for a given model,
should have errors well below the observational error, which in the
case of the CoRoT mission is expected to be below 0.1\,$\mu$Hz
\citep{baglin}.  For the Kepler mission \citep[e.g.,][]{Christ2007},
with expected launch in early 2009, selected stars may be observed
continuously for several years and errors as low as $10^{-3} \, \mu$Hz
may be reachable, particularly for modes excited by the heat-engine
mechanism.  Since errors resulting from numerical problems are
typically highly systematic, they may affect the asteroseismic
inferences even if they are substantially below the random errors in
the observed frequencies.  This must be kept in mind in the assessment
of the estimates of the numerical errors.

During the last decades a lot of codes obtaining numerical solutions
of an adiabatic system of differential equations describing stellar
oscillations have been developed. In order to ascertain whether any
possible difference in the description of the same observational data
by different numerical codes is due to physical descriptions or to
different numerical integration schemes, the inter-comparison of these
oscillation codes in a fixed and homogeneous framework is absolutely
necessary. Some effort has been already done in the past but only
regarding pairs of codes. Some codes have also developed a lot of
internal precision tests. However, there is a lack of inter-comparison
of a large enough set of codes. We aim in this study try to fix a set
of minimum requirements for a code to be sure that any difference
found is only due to a different physical assumption.

Ideally, for a given model there should be a set of `true' frequencies
with which the results of the different codes could be compared.  This
ideal situation could probably be approximated by considering
polytropic models for which it is relatively straightforward to
calculate the equilibrium structure with essentially arbitrary
accuracy \citep[see also][]{Christ1994}.  In practice, the situation
for realistic stellar models is more complex.  Owing to the complexity
of the stellar evolution calculation the models are often available on
a numerical mesh which is not obviously adequate for the pulsation
calculation.  The effect of this on the frequency computation depends
on the detailed formulation of the equations in the pulsation codes.
These formulations are equivalent only if the equilibrium model
satisfies the `dynamical' equations of stellar structure, i.e., the
mass equation and the equation of hydrostatic support, and this is
obviously not exactly true for a model computed on a finite (possibly
even relatively sparse) mesh.  One might define a consistent set of
frequencies for a given model by interpolating it onto a very dense
mesh and resetting it to ensure that the relevant equations of stellar
structure are satisfied.  The model is fully characterized by the
dependence on distance $r$ to the centre of density $\rho$ and the
adiabatic exponent $\Gamma_1 = (\partial \ln P/ \partial \ln
\rho)_{\rm ad}$, $P$ being pressure and the derivative being at
constant specific entropy.  Thus one could interpolate $\rho(r)$ and
$\Gamma_1(r)$ to a fine mesh, and recompute the mass distribution and
pressure by integrating the mass equation and the equation of
hydrostatic equilibrium.  Frequencies of this model should then be
essentially independent of the formulation of the oscillation equation
and would provide a suitable reference with which to compare other
frequency calculations.  Such a test may be carried out at a later
stage in the ESTA effort.

In Task~2, for now, we aim at testing, comparing and optimizing the
seismic codes used to calculate the oscillations of existing models of
different types of stars. In order to do so we consider steps in the
comparison by addressing some of the most relevant items that must be
compared regarding the seismic characterization of the models:

\begin{itemize}
    \item Step 1: comparison of the frequencies from different seismic
    codes for the same model.
    \item Step 2: comparison of the frequencies from the same seismic
    code for different models of the same stellar case provided by
    different equilibrium codes.
    \item Step 3: comparison of the frequencies for specific
    pulsators.
\end{itemize}

The work presented here is mostly focused on step~1. At this step
three different equilibrium models have been used. Two of them have
been computed using \CESAM\ \citep{cesam}, %(Morel and Lebreton, this volume)
with 902 and 2172 mesh points, and a third one with 4042 mesh points provided
by \ASTEC\ \citep{astec}.% (Christensen-Dalsgaard, this volume).
We present
inter-comparisons using the two models with the larger numbers of mesh
points. The same physics and physical constants (except $G$) are used
for all the oscillation codes. Frequencies in the range of $[20,2500]
\,\mu$Hz, belonging to spherical degrees $\ell=0,1,2$ and 3 have been
calculated, in order to recover most of the possible values we can
find in the observational photometric data.

We present in Sect. 2 the equilibrium models used and we analyze their
main features. Section 3 is devoted to the different oscillation codes
used,   discussing   common  requirements   and   variations  in   the
treatment.  In  Sect.  4  the  direct comparison  of  frequencies  for
different  frequency  ranges and  spherical  degrees  is presented  in
detail. Sections  5 and  6 analyze the  same inter-comparison  for the
values of  the large and  small separations, respectively.   Section 7
discusses the  dominant effects that contribute to  the differences in
frequencies determined  by the different codes. Conclusions are given
in Sect. 8.

%shows the differences encountered
%when different values of the gravitational constant $G$ found recently
%in the literature are used.

%-------------------------------------------------------------------
\section{The equilibrium models.}\label{sec:1}

To ensure that any difference obtained in the inter-compari\-sons is
only due to differences in the numerical schemes, we imposed to all
the codes the use of the same equilibrium models. These models were
supplied in several formats: {\tt OSC}, {\tt FGONG}, {\tt SROX}, and
{\tt FAMDL}. The first model was required to have 900 mesh points, and
it was provided by \CESAM. The differences in this case reached
unacceptable values of 2-3 $\mu$Hz when the same integration schemes
are compared, and even 10 $\mu$Hz when the use or not of the
Richardson extrapolation are compared. The maximum difference found for
large separations with this model is $1 \,\mu$Hz. This showed that
either a larger number of mesh points or Richardson extrapolation
was necessary.%
\footnote{Detailed results of this investigation can be found at
\hfill\break
{\tt http://www.astro.up.pt/corot/compfreqs/task2/}.}

\begin{table}[t]
\caption{General characteristics of the models used for the
  inter-comparison.}
\centering
\label{tab:1}       % Give a unique label
\begin{tabular}{lllllll}
\hline\noalign{\smallskip}
$M/M_\odot$ & $\log T_{\rm eff}$ & Age & $X_{\rm c}$ & $R/R_\odot$
& Mesh & $G$\\
 &  &(in My) &  &  & points & $[10^{-8}\, {\rm cgs}]$\\[3pt]
\tableheadseprule\noalign{\smallskip}
1.5 & 3.826 & 1355 & 0.4 & 1.731 & 4042 & 6.6716823\\
1.5 & 3.830 & 1368 & 0.4 & 1.724 & 2172 & 6.67232\\
\noalign{\smallskip}\hline
\end{tabular}
\end{table}
%==========================

\begin{figure}
\centering
%\begin{tabular}{cc}
  \rotatebox{-90}{\includegraphics[width=6cm]{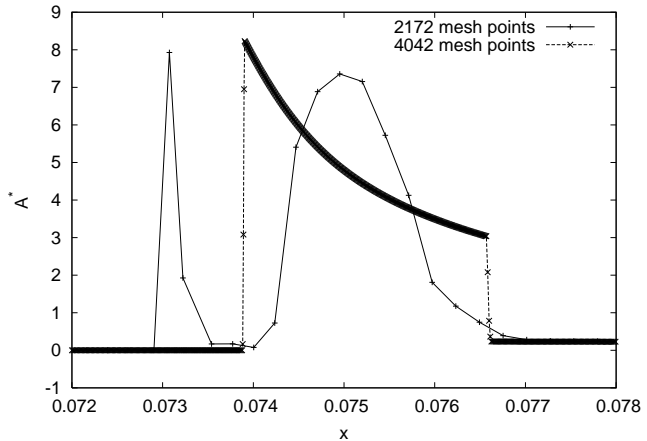}}
%\end{tabular}
\caption{$A^*$ (related to the Brunt-V\"ais\"al\"a frequency) as a
  function of the relative radius for the two equilibrium models
  discussed in the text in the $\mu$-gradient zone, close to the
  convective core.
  The mesh points provided in the models are indicated.}
\label{fig:1}       % Give a unique label
\end{figure}

\begin{figure}
\centering
%\begin{tabular}{cc}
  \rotatebox{-90}{\includegraphics[width=6cm]{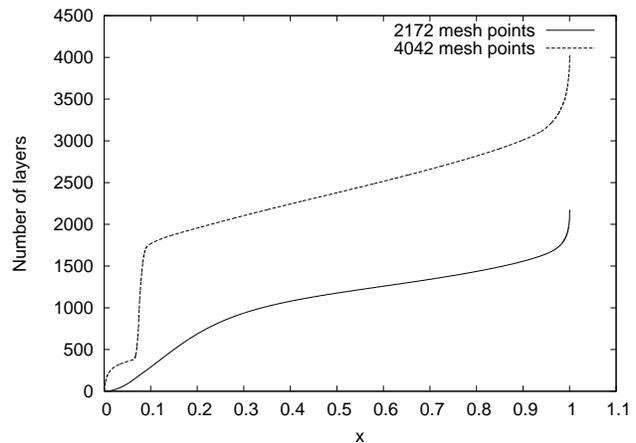}}
%\end{tabular}
\caption{Accumulated number of layers as a function of the relative
  radius for the two equilibrium models explained in the text.}
\label{fig:2}       % Give a unique label
\end{figure}

The second model was also provided by \CESAM\ (referred from now on as
M2k). It uses a grid, with 2172 mesh points, more suitable for
asteroseismic purposes. General characteristics of the model are
presented in Table 1. These are typical of a $\gamma$ Doradus star
showing oscillations in the asymptotic g-mode regime, and also around
the fundamental radial mode. A priori, solar-like pulsations cannot be
excluded for this type of star and therefore it can be a good
candidate for a global study. In Fig. \ref{fig:1}, $A^*$ (which is a
quantity directly related to the Brunt-V\"ais\"al\"a frequency $N^2$:
$A^*=rg^{-1}N^2$, where $g$ is the gravitational acceleration) is
depicted as a function of the relative radius ($x=r/R$) in a region of
steeply varying hydrogen abundance, and hence mean molecular weight
$\mu$, just outside the convective core. The model is in a phase of a
growing convective core.  If diffusion and settling are neglected this
leads to a discontinuity in the hydrogen abundance and hence,
formally, to a delta function in $A^*$; also, there is a region of
`semiconvection' at the edge of the core.  In fact, the figure shows
that three points in this model display an erratic variation in $A^*$
just in the transition region between the convective and the radiative
zone; also, the mesh resolution of this region of rapid variation
seems inadequate.  This is emphasized by Fig. \ref{fig:2} which shows
the distribution of mesh points of this model along the stellar
radius, indicating that there are not enough mesh points in this
transition zone.  As discussed below, these features of model M2k give
rise to frequency differences in the comparison, particularly for
those modes for which this inner part is critical for their physical
description.

The third and last model is a 4042 mesh points model (from now on M4k)
provided by \ASTEC. General characteristics of the model are
presented in Table 1. It has been computed to have overall
characteristics similar to the previous \CESAM\ model in order
to understand better the differences. 
However, as discussed by Christensen-Dalsgaard (this volume) particular
care has been taken in the treatment of the $\mu$-gradient region;
the semiconvective region was replaced by a region with a steep gradient
in the hydrogen abundance, defined such as to ensure neutral stability
of the temperature gradient.
As a result, $A^*$ for this model, also 
presented in Fig. \ref{fig:1}, shows a well-defined and reasonable behaviour.
The distribution of the mesh points can be
found in Fig. \ref{fig:2}. 
In the central and outer parts of the stellar model, the distribution
is similar to the \CESAM\ model M2k. It is in the inner zones,
particularly in the
boundary region between the convective core and the $\mu$-gradient
zone, that the models present different mesh-point
distributions, with M4k providing a far superior resolution of this
critical region.

%-------------------------------------------------------------------
\section{Oscillation codes and requirements.}

All oscillation codes involved in this task were asked to provide
adiabatic frequencies in the range of $[20,2500] \,\mu$Hz and spherical
degrees $\ell=0,1,2$ and 3. In addition, the solution of the equations
must satisfy the following requirements:

\begin{table*}[t]
\caption{List of participating codes in this inter-comparison and
  ``degrees of freedom" for each code: (a) Eigenfunctions, (b) Order of
  the Integration Scheme (2 or 4), (c) Use (y) or not (n) of the
  Richardson extrapolation, (d) integration variable used,
  and (e) the choice of the gravitational constant $G$. The
  references for each code are also given.}
\centering
\label{tab:2}       % Give a unique label
\begin{tabular}{lllllll}
\hline\noalign{\smallskip}
Code & E.F. & I.S. & Rich. & I.V. & $G$             & Reference\\
     &      &      &       &      & $[10^{-8}\, {\rm cgs}]$ &\\[3pt]
\tableheadseprule\noalign{\smallskip}
\ADIPLS & $P^\prime$ & 2 & y, n & $r$ & 6.67232 & \citet{adipls} \\
\FILOU\  & $P^\prime$ & 2 & n & $r$ & 6.67232 & \citet{filou} \\ 
\GRACO\  & $P^\prime$ , $\delta P$ & 2 & y, n & $r$, $r/P$ & 6.6716823 & \citet{graco} \\ 
\LOSC\ & $\delta P$ & 4 & n & $r$ & 6.67232 & \citet{losc} \\
\NOSC\ & $P^\prime$ , $\delta P$ & 2 & y, n & $r$, $r/P$ & 6.67259 & \citet{nosc} \\
\OSCROX\ & $P^\prime$ & 4 & n & $r$ & 6.6716823 & \citet{oscrox}\\
\POSC\ & $P^\prime$ & 2 & y, n & $r$ & 6.6716823 & \citet{posc}\\
\PULSE\ & $P^\prime$ & 4 & n & $r/P$ & 6.6716823 & \citet{pulse}\\
\ROMOSC\  & $P^\prime$ & 2 & n & $r$ & 6.67232 & \citet{romosc}\\
\noalign{\smallskip}\hline
\end{tabular}
\end{table*}

\begin{itemize}

\item To use the mesh provided by the equilibrium model, no re-meshing
  is allowed.

\item To set the Lagrangian perturbation to the pressure to zero ($\delta
  P=0$) as outer mechanical boundary condition.

\item To use the physical constants prescribed in Task~1.

\item To use linear adiabatic equations.

\end{itemize}

Nevertheless, some other schemes for the numerical solution of the
differential equations (from now on called for simplicity ``degrees of
freedom'') remain open.
Nine oscillation codes of different research groups in the field have
been used in this inter-comparison exercise. A summary of the
participating codes and the different ``degrees of freedom" provided by
each one is found in Table~2 and include:

\begin{itemize}

\item Set of eigenfunctions: Use of the Lagrangian or the Eulerian
  perturbation to the pressure ($\delta P$ or $P^\prime$). 
  This obviously affects the form of the equations; in particular,
  when using $\delta P$ the equations do not depend explicitly on $A^*$.

\item Order of the integration scheme: Most of the codes use a
  second-order scheme, but some of them have implemented a
  fourth-order scheme.

\item Richardson extrapolation: Some of the codes using a se\-cond-order
  scheme have the possibility to use Richardson extrapolation
  \citep{shiba} to decrease the truncation error; combining 
  a second-order scheme with Richardson extrapolation yields errors
  scaling as ${\cal N}^{-4}$, ${\cal N}$ being the number of mesh points
  \citep[e.g.][]{Christ1994}.

\item Integration variable: Two integration variables are used: 1) the
  radius ($r$), or 2) the ratio $r/P$. The latter variable may reduce the
  effect of rounding errors in the outer layers 
(see Sect. \ref{sec:indep}).

\item Despite the requirement that the physical constants be fixed at the
  values for Task~1, the codes used slightly different values 
  of the gravitational constant $G$, as listed in Table~\ref{tab:2}.
  Ideally the equilibrium model should have been computed with the prescribed
  Task~1 value ($6.6716823 \times 10^{-8}$\,cgs) which should then have been
  used for the oscillation calculations.
  In practice Model M4k was computed with $G = 6.67232 \times 10^{-8}$\,cgs.
  Using different values of $G$ in the oscillation equations clearly gives
  rise to inconsistencies, with potential effects on the frequencies,
  as discussed further in Sect.~\ref{sec:grav}.

\end{itemize}

Note that most of the oscillation codes put $\ell=0$ in the general
non-radial differential equations to obtain the radial modes, except
for \LOSC\ that uses the LAWE differential equation (Linear Adiabatic
Wave Equation for radial modes), and for \GRACO\ for which results
will be shown for both sets of equations.  It should be noted that the
LAWE does not depend on $A^*$.  All the main characteristics and
numerical schemes are presented in previous chapters of this volume.

The Nice code (\NOSC) has the options of using $P'$ or $\delta P$ as
dependent variables, and $r$ or $r/P$ as independent variable.
However, all \NOSC\ results presented here use $P'$ and $r$.

The use of the mesh provided in the equilibrium model, rather than meshes
optimized for the different kinds of modes, may result in inadequate resolution
of the rapid variation of high-order p- and g-modes and hence larger truncation
errors in the solution of the differential equations.
In these cases, therefore, the second-order scheme may result in 
unacceptable errors. 
Here the use of higher-order integration schemes (a fourth-order scheme
or a second-order scheme followed by Richardson extrapolation)
is therefore expected to give better results.
For low- and intermediate-order modes we expect little effect of the
use of higher-order schemes.

%-------------------------------------------------------------------
\section{Frequency inter-comparisons.}

\begin{figure}
\centering
%\begin{tabular}{cc}
  \rotatebox{-90}{\includegraphics[width=6cm]{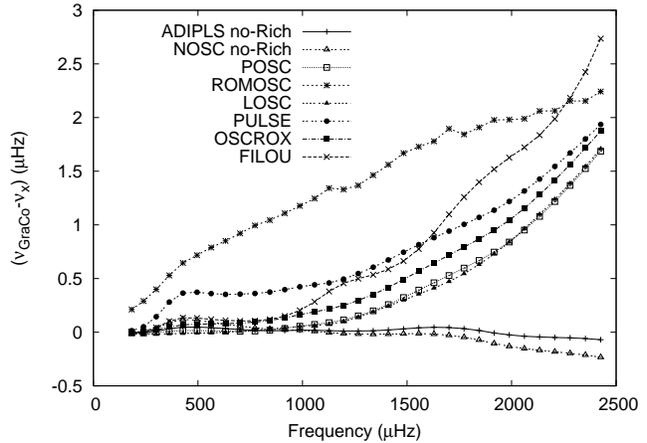}}
%\end{tabular}
\caption{Frequency comparison (reference line is \GRACO) for modes with
   $\ell=0$ obtained for the model M2k. \ADIPLS\ and \NOSC\ frequencies
   have been obtained without using the Richardson extrapolation.}
\label{fig:3}       % Give a unique label
\end{figure}

In this section, the results of the direct frequency
inter-com\-pa\-rison are presented. We have structured this study
splitting the frequency range in three parts for the non-radial case,
only one for the radial case, and comparing codes only with similar
selections of ``degrees of freedom". In addition, the influence of using
different selections in the same code (\GRACO\ in this case) is shown;
for these tests the value of $G$ in the
oscillation calculations was the same as was used to compute the equilibrium
model.

%..............................................
\subsection{Radial modes.}

These are shown in a single frequency range. In Fig. \ref{fig:3} the
results obtained using the model M2k are presented. The reference line
for all the inter-com\-pa\-risons is selected to be \GRACO. In this
figure the reference frequencies have been obtained using $P^\prime$,
second order, no Richardson extrapolation and $r$ as independent variable (see
Table~2).  Two sets of codes can be identified in the figure,
\ADIPLS-\NOSC-\GRACO, with differences lower than $0.25 \,\mu$Hz, and all
with the same ``degrees of freedom", and \OSCROX-\PULSE-\LOSC-\POSC\ with
differences for high frequencies lower than $2 \,\mu$Hz with \GRACO,
but with differences among them around $0.5 \,\mu$Hz. This second set
of codes differs from the first one in the use of a fourth-order
numerical scheme instead of a second-order one. 
In the present figure, in Fig.~\ref{fig:5} and in Figs \ref{fig:7} 
and \ref{fig:8}, showing inter-comparisons with model M2k,
\POSC\ has been chosen as
representative of the codes using second order plus Richardson extrapolation
(see Table~\ref{tab:2}).
%\note [I changed the `From now on' since these are now the only relevant
%cases; I hope that this clears up one of the Li\`ege questions!].
We can see how the use of this integration procedure
provides similar results as the fourth-order solutions. These
differences, around $0.5 \,\mu$Hz for codes using the same integration
scheme, and $2 \,\mu$Hz for codes using different schemes, are larger
than the expected precision of the coming observational
data. Therefore this effect can change any detailed physical
description as interpreted by different oscillation codes. Also,
we point out that the differences between the codes using
fourth-order schemes and \GRACO\ results using a second-order scheme
indicate that the model has an insufficient number
of mesh points for asteroseismic studies.

\begin{figure}
\centering
%\begin{tabular}{cc}
  \rotatebox{-90}{\includegraphics[width=6cm]{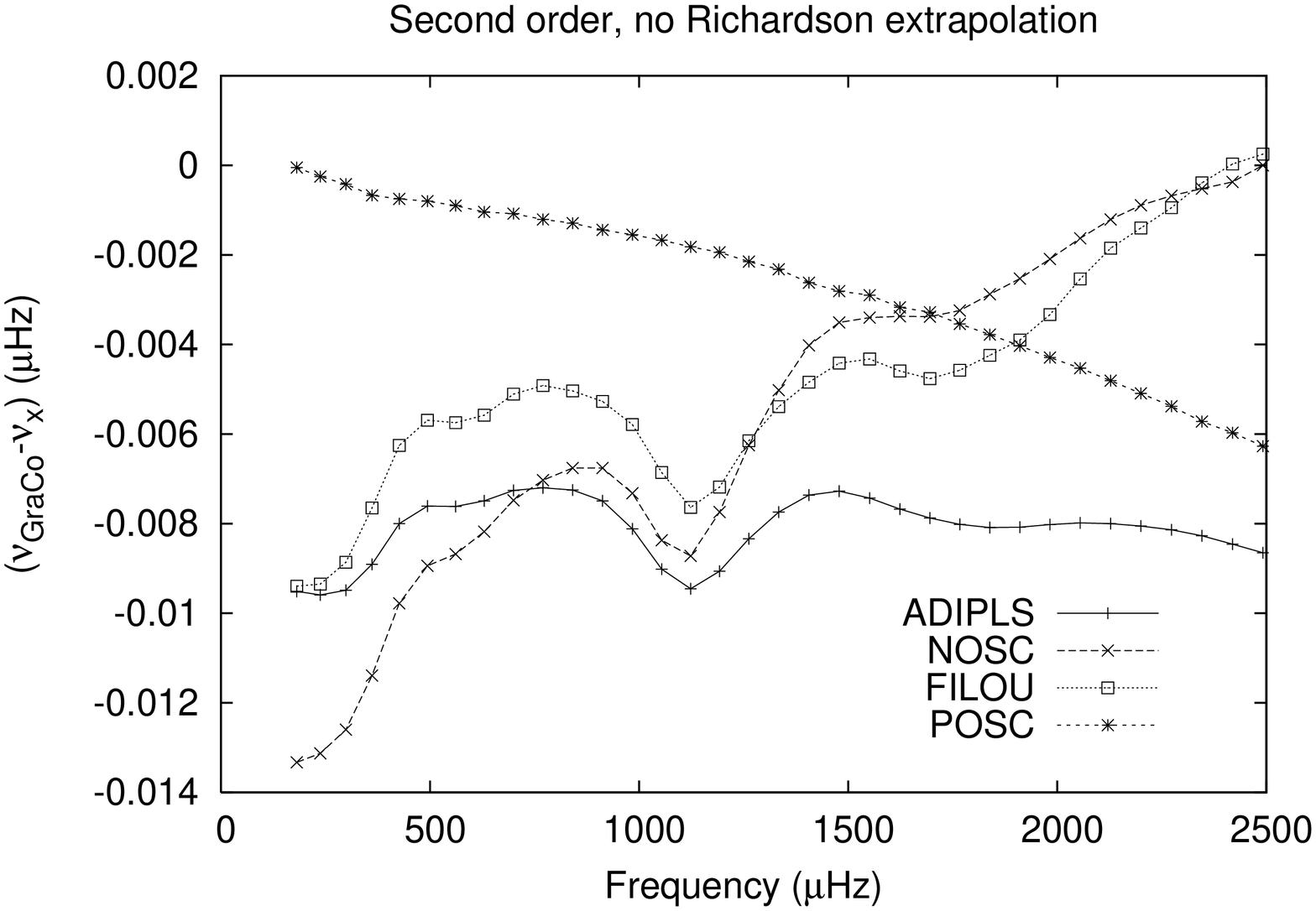}}
  \rotatebox{-90}{\includegraphics[width=6cm]{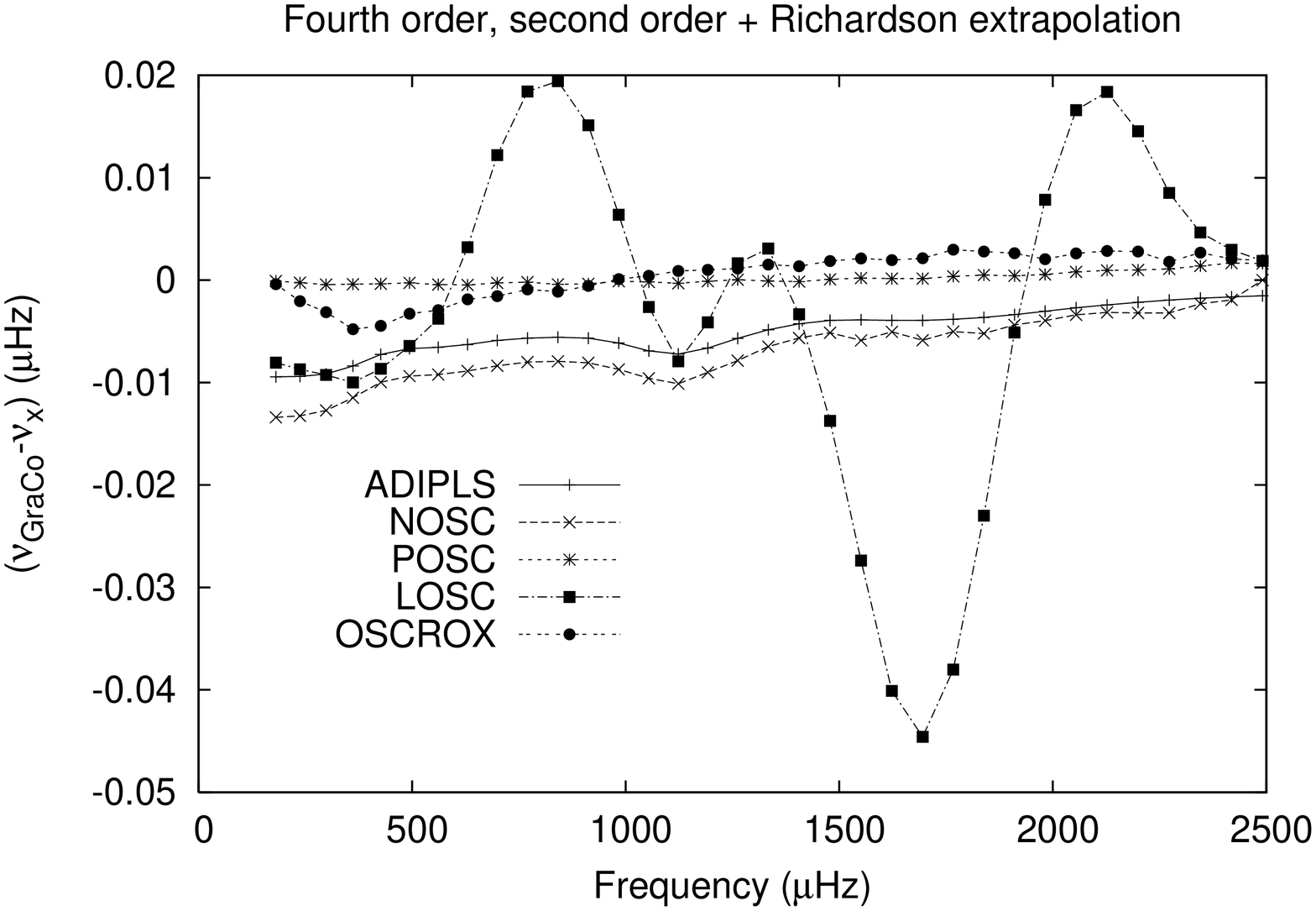}}
  \rotatebox{-90}{\includegraphics[width=6cm]{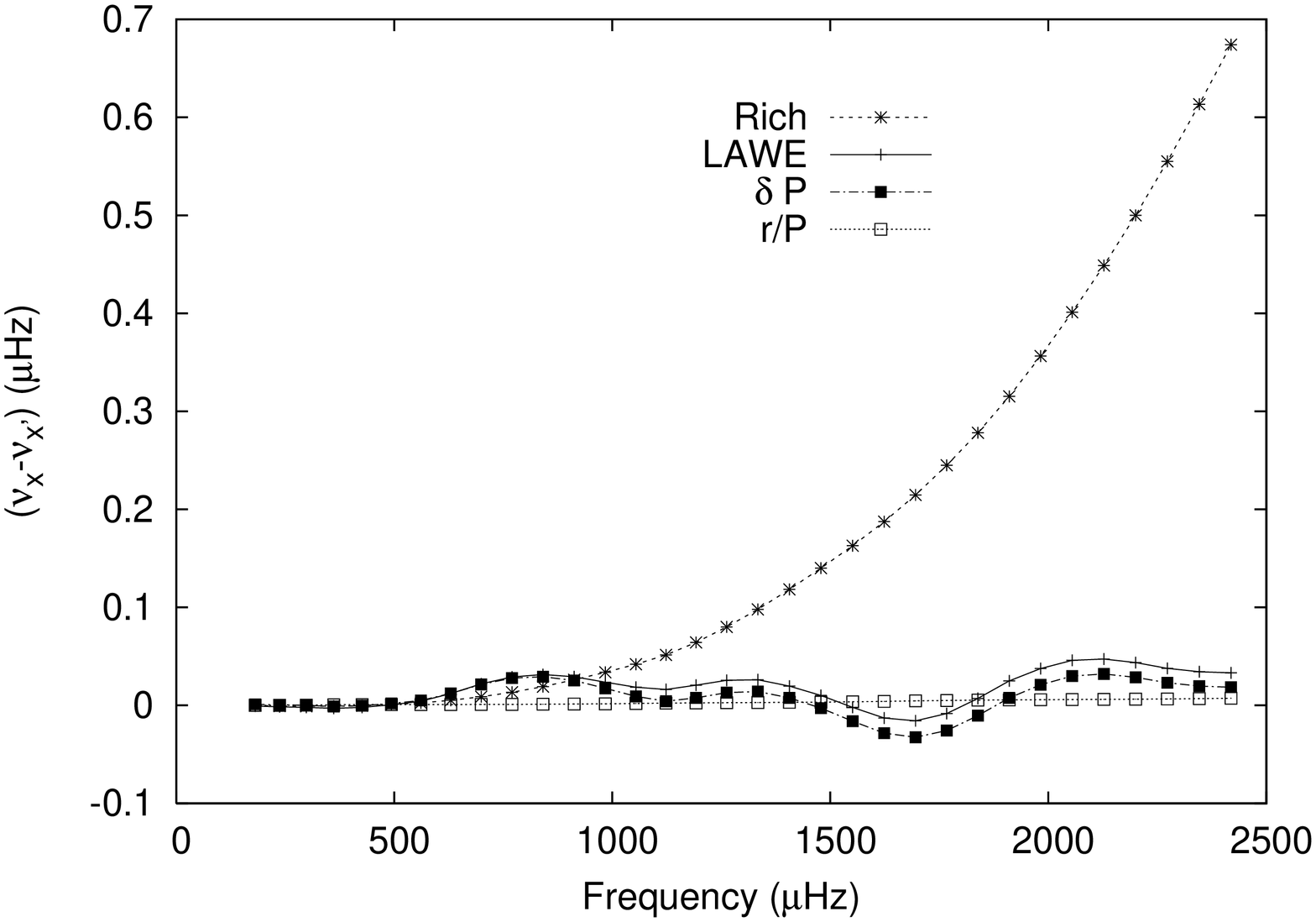}}
%\end{tabular}
\caption{Frequency comparison (reference line is \GRACO) for modes with
   $\ell=0$ as a function of the frequency obtained for
   the model M4k. In the top panel the models with: second order, no
   Richardson extrapolation,
   $r$, are depicted (\NOSC-\ADIPLS-\GRACO-\FILOU-\POSC). The
   middle panel presents the differences obtained for models providing
   fourth-order integration solutions or second-order plus
   Richardson extrapolation (\LOSC-\OSCROX-\NOSC-\ADIPLS-\GRACO-\POSC). In
   the bottom panel a comparison of the different ``degrees of freedom"
   only using \GRACO\ is presented.}
\label{fig:4}       % Give a unique label
\end{figure}

Figure~\ref{fig:4} shows the results obtained for the model M4k. In the
top panel models with second order, no Richardson extrapolation,
$r$, are depicted. 
All the differences are lower in magnitude than $0.014 \,\mu$Hz,
i.e., two orders of magnitude lower than those obtained for
model M2k. Therefore improving the mesh, including
a doubling of the number of points,
provides a very substantial improvement in the precision, making these
values more acceptable for theoretical modeling.

The middle panel of Fig. \ref{fig:4} presents the differences obtained
with models providing fourth-order integration solutions or
second-order plus Richardson extrapolation.
The global precision here is similar to the previous case or even 
slightly lower, with differences lower than $0.02 \,\mu$Hz. It is interesting
to point out that we cannot directly distinguish between a
fourth-order integration scheme solution (\OSCROX) or a
second-order plus Richardson extrapolation (the rest).
However, it is noticeable that the \GRACO-\OSCROX-\POSC\ fall in one group and
\ADIPLS-\NOSC\ in a second, with a slight difference in the latter case.
These two groups are distinguished by the value of $G$
(cf.\ Table~\ref{tab:2}),
with \ADIPLS\ and \NOSC\ having similar but not identical values.
This pattern will be found in other cases also.
The \LOSC\ behaviour is discussed in the next paragraph.

Finally in the bottom panel of Fig. \ref{fig:4} a comparison between
the different ``degrees of freedom" using only the \GRACO\ code are
presented. As reference we have used the solution with ``degrees of
freedom": $X=$($\ell=0$, no Richardson extrapolation, $P^\prime$,
$r$). For each comparison we have changed only one of these ``degrees
of freedom" at a time, keeping the rest unchanged (solutions $X'$).
The most prominent effect arises from the use of Richardson
extrapolation, which changes the frequencies by nearly $0.8 \,\mu$Hz
for the highest-order modes, substantially more than the expected
observational accuracies. For model M2k (see Fig. \ref{fig:3}), we
have similarly found a change of $2 \,\mu$Hz, reflecting the smaller
number of mesh points.  This clearly shows that second-order schemes
are inadequate, even for the mesh in M4k, for the computation of
high-order acoustic modes; as expected the effect decreases rapidly
with decreasing mode order.  The use of $r/P$ as integration variable
provides small differences, always lower than $0.008\, \mu$Hz. These
differences are of the order of those obtained in the top and middle
panels.  The use of the Lagrangian perturbation to the pressure as
variable ($\delta P$) or the use of the LAWE differential equation
provide very similar differences, lower than $0.05 \,\mu$Hz, but with
an oscillatory pattern. This pattern is very similar to that observed
for \LOSC, which also uses LAWE to obtain the radial modes.  As
discussed in Sect.~\ref{sec:dep}, this oscillatory pattern arises from
an inconsistency in the thermodynamics of model M4k which affects
$A^*$; solutions of equations that do not depend on $A^*$ (i.e., the
LAWE or the equations based on $\delta P$) are insensitive to this
effect.  Therefore, even for model M4k and radial modes, the use of
different integration procedures can give different values for the
oscillation eigenfrequencies that are non-physical in nature. These
non-physical sources of differences are mainly some inconsistencies in
the equilibrium models (see Sect.~\ref{sec:dep}) and the lack of mesh
points. But when the same numerical schemes are used, the different
codes provide very similar frequencies.

%..............................................
\subsection{Non-radial modes with $\ell=2$}

To illustrate the differences appearing in the case of non-radial
modes, the spherical degree $\ell=2$ has been chosen. We have divided
the frequency spectrum into three regions: 1) Large frequencies
($[500,2500] \,\mu$Hz), 2) frequencies a\-round the fundamental radial
mode ([80,500] $\mu$Hz),
and 3) low-frequency region ($[20,80] \,\mu$Hz).
In all cases a study similar to that developed in the
radial case has been carried out.

For the sake of simplicity the high-frequency differences are not
represented since the results are very similar to those presented for
the radial case. Only a slightly higher precision is found in this
case. The results of \LOSC\ present the same pattern as the radial case,
owing to the use of $\delta P$ as eigenfunction in that code (see
Sect. ~\ref{sec:dep}).

\begin{figure}
\centering
%\begin{tabular}{cc}
  \rotatebox{-90}{\includegraphics[width=6cm]{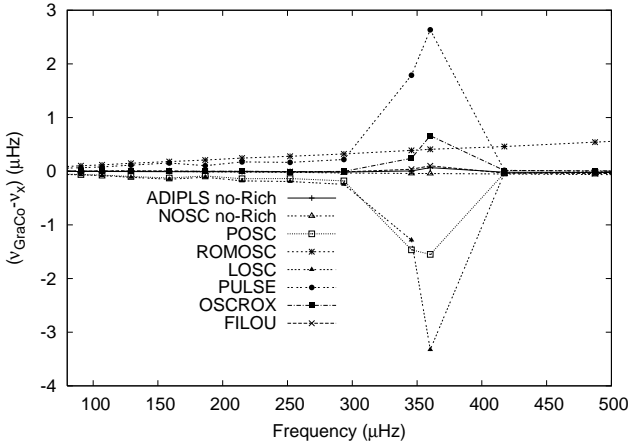}}
%\end{tabular}
\caption{Frequency comparison (reference line is \GRACO) for modes with
   $\ell=2$ for the model M2k around the
   fundamental radial mode. \ADIPLS\ and \NOSC\ frequencies have been
   obtained without using Richardson extrapolation.}
\label{fig:5}       % Give a unique label
\end{figure}

Figure \ref{fig:5} shows the results obtained for model M2k when
comparing $\ell=2$ frequencies around the fundamental radial mode. The
main differences are smaller than in the high-frequency
region, corresponding to the low order of the modes and the
consequent lesser sensitivity to the number of mesh points.
The largest differences are found for two modes
of frequency near $345$ and $362 \, \mu$Hz showing avoided crossing;
these modes have a mixed character, with fairly substantial amplitude
in the $\mu$-gradient zone.
\PULSE\ and \LOSC\ present differences around
$3-4\,\mu $Hz, \POSC\ around $2\,\mu $Hz, \OSCROX\ less than $1\,\mu $Hz,
and the rest do not present significant differences for these two
modes. That is, the largest differences are found in these codes when
using a fourth-order integration scheme or a second-order plus
Richardson extrapolation. The values of these differences are larger than
the expected precision of the satellite data to come.
They clearly reflect the inadequate representation of the 
$\mu$-gradient zone in M2k,
with higher-order schemes being more sensitive to the resulting
inconsistency in the model.

\begin{figure}
\centering
%\begin{tabular}{cc}
  \rotatebox{-90}{\includegraphics[width=6cm]{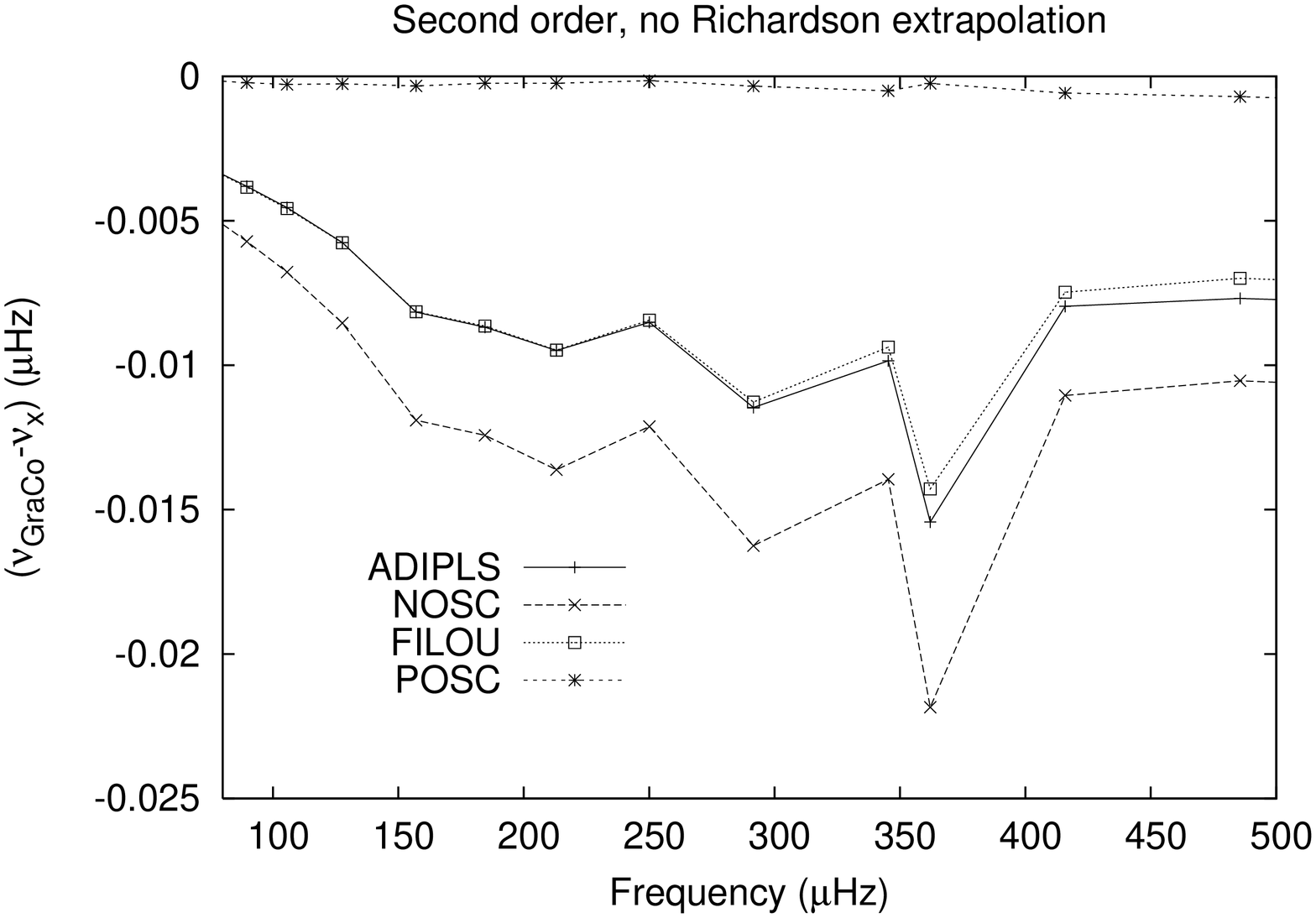}}
  \rotatebox{-90}{\includegraphics[width=6cm]{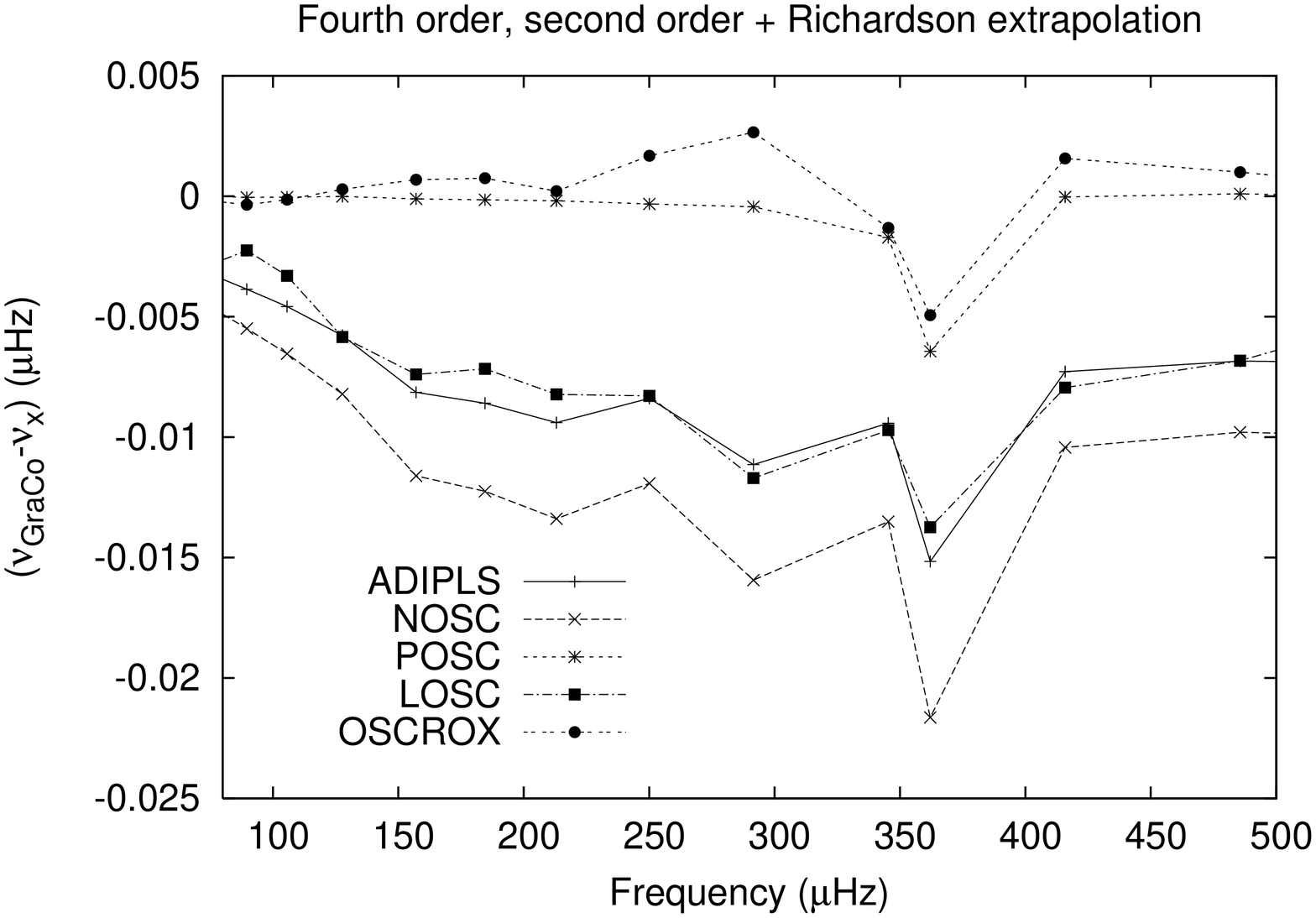}}
  \rotatebox{-90}{\includegraphics[width=6cm]{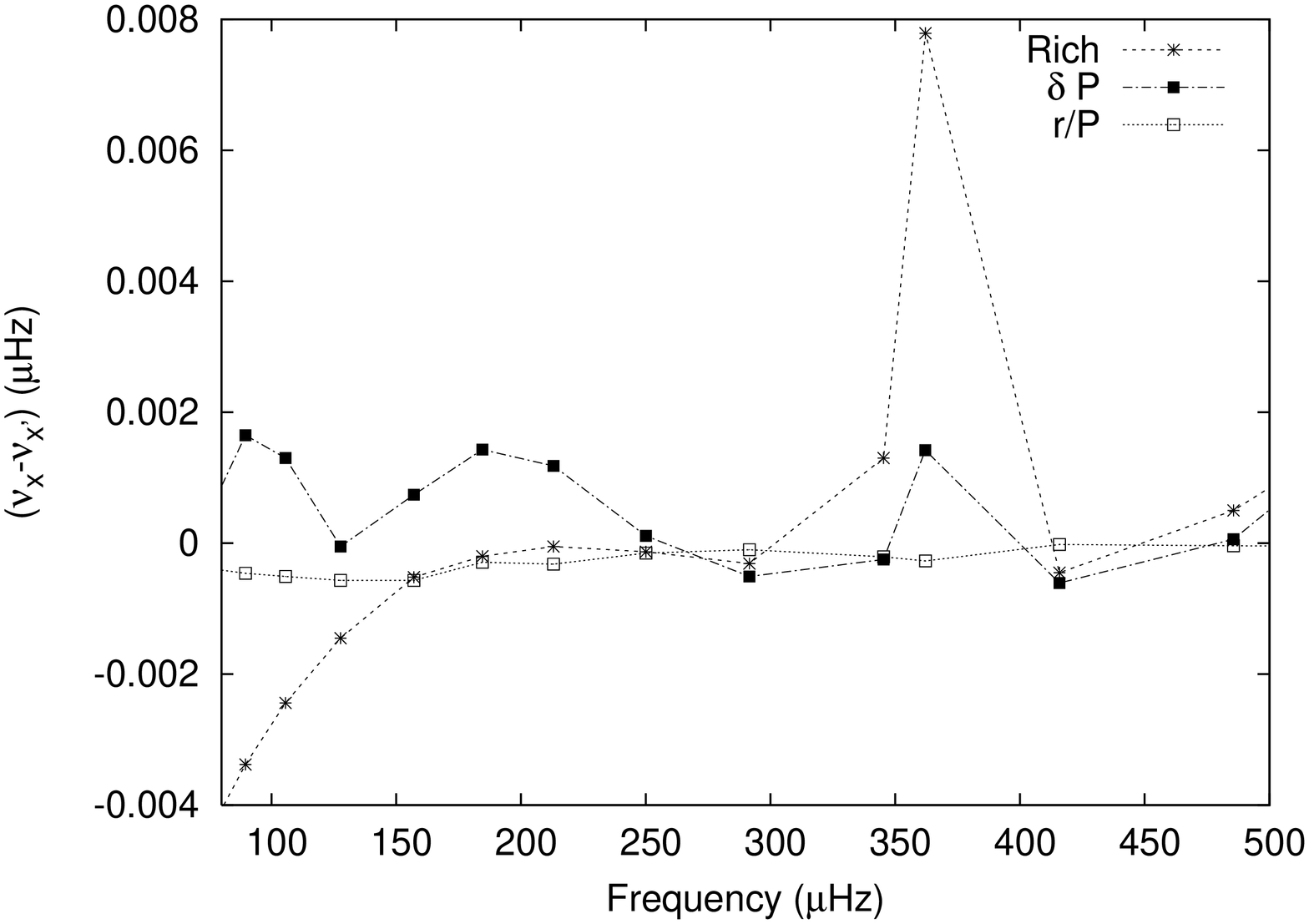}}
%\end{tabular}
\caption{Frequency inter-comparison (reference line is \GRACO) for modes with
   $\ell=2$ as a function of the frequency obtained for the
   model M4k and for modes around the fundamental radial. In top
   panel models with: second order, no Richardson extrapolation, and $r$ are
   depicted (\NOSC-\ADIPLS-\GRACO-\FILOU-\POSC). Middle panel presents the
   differences obtained for the models providing fourth-order
   integration solutions or second-order plus Richardson
   extrapolation (\LOSC-\OSCROX-\NOSC-\ADIPLS-\GRACO-\POSC). In bottom panel
   an inter-comparison for the ``different degrees" of freedom only with
   \GRACO\ is presented.}
\label{fig:6}       % Give a unique label
\end{figure}

In the top panel of Fig. \ref{fig:6} the inter-comparisons of the
frequencies ($\ell=2$) for model M4k, using a second-order scheme without 
Richardson extrapolation, are depicted for the same frequency range as in
Fig. \ref{fig:5}. The main differences are two orders of magnitude
lower than those obtained for model M2k, and they are also of the
same order of magnitude as those obtained for the high-frequency range
with M4k.  All these differences remain always lower than $0.025 \,\mu$Hz. 
\GRACO\ and \POSC\ are extremely close, while 
\FILOU\ and \ADIPLS\ provide very similar results, with slightly larger
differences for \NOSC, by
about $0.005 \,\mu$Hz, relative to these codes.
Thus again the differences are related directly to the different values of $G$.
There remain small wiggles for the
mixed modes near $350 \, \mu$Hz but reduced more than two orders of
magnitude relative to the largest differences for model M2k,
reflecting the superior resolution of the critical region in model M4k.

The middle panel of Fig. \ref{fig:6} shows the same inter-com\-pa\-rison
as the top panel for codes using a fourth-order scheme or a
second-order plus Richardson extrapolation. The precision is similar to
the previous case, with differences of the same order of
magnitude. We can distinguish two groups of codes providing very
similar results: \OSCROX-\POSC-\GRACO\ and \LOSC-\ADIPLS-\NOSC. This
distribution does not depend on the integration scheme, but
again reflects the values of $G$.
%\note [likely -- we still need the value for \LOSC. !jcd]
The wiggle of
the mixed modes is also very similar to that obtained with
a second-order scheme without Richardson extrapolation.

\begin{figure}
\centering
%\begin{tabular}{cc}
  \rotatebox{-90}{\includegraphics[width=6cm]{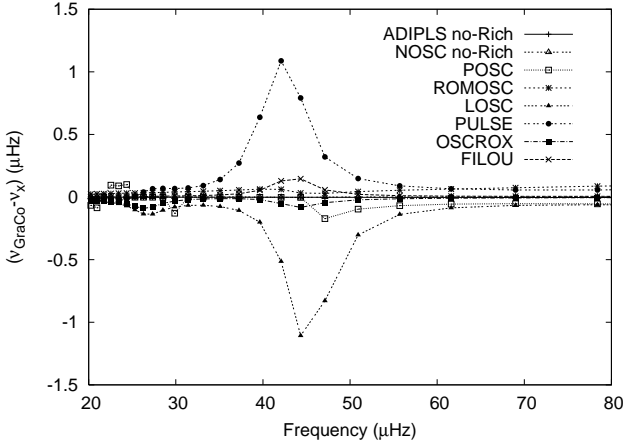}}
%\end{tabular}
\caption{Frequency comparison (reference line is \GRACO) for modes with
   $\ell=2$ obtained for the model M2k in the low
   frequency region. \ADIPLS\ and \NOSC\ frequencies have been obtained
   without using the Richardson extrapolation.}
\label{fig:7}       % Give a unique label
\end{figure}

\begin{figure}
\centering
%\begin{tabular}{cc}
  \rotatebox{-90}{\includegraphics[width=6cm]{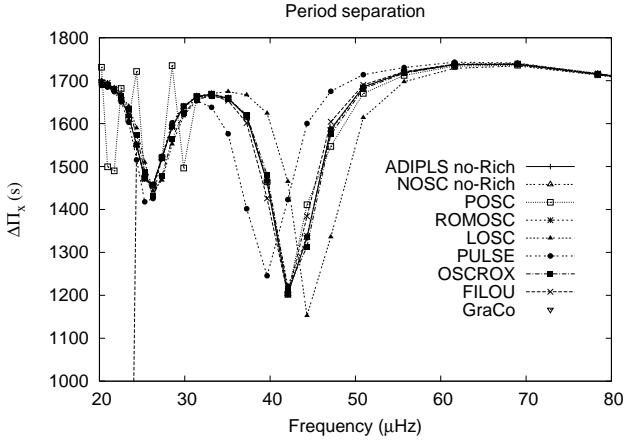}}
%\end{tabular}
\caption{Period separation as a function of the frequency in the
   asymptotic g-mode region for model M2k.}
\label{fig:8}       % Give a unique label
\end{figure}

Finally, an inter-comparison of different ``degrees of freedom" using only
\GRACO\ is presented in order to test the differences obtained for
the different choices. This comparison is depicted in the bottom panel
of Fig. \ref{fig:6}. 
The use of the Richardson extrapolation is not very important,
as expected for these modes of low order,
with effects generally smaller than $0.002 \,\mu$Hz,
although a larger value is present for the mixed modes. 
Using $\delta P$ as variable
gives differences larger than the differences among
codes with the same ``degrees of freedom". The use of the integration
variable $r/P$ does not introduce significant differences.
%\note [As noted in a mail, the result using $\delta P$
%seems inconsistent with the \ADIPLS/\LOSC\ comparison in the middle panel. 
%It is difficult to understand that this should be a result of using 
%a second-order scheme. I hope that this can be understood better!]
%That is, in this frequency
%spectrum region, the integration variable is the only significant
%source of differences due to the numerical integration schemes.

Compared with model M2k, we find a general reduction of the
differences for M4k, the main effects being in the region of mixed
modes with an improvement reaching up to three orders of magnitude.
This is obviously not a simple consequence of the doubling of the
number of mesh points.  The main reason is likely the inadequate
resolution shown by model M2k in the description of the
Brunt-V\"ais\"al\"a frequency in the region close to the boundary of
the convective core, which is not present in model M4k. The
avoided-crossing phenomenon and the behaviour of the mixed modes are
very sensitive to the detailed treatment of this region,
including the effects of semiconvection.
Therefore,
an accurate description of $N^2$ is critical for the oscillation codes
in order correctly to obtain eigenfrequencies for modes near an
avoided crossing.

The direct frequency inter-comparison ends with the stu\-dy of the
low-frequency region;
as above we concentrate on modes of degree $\ell = 2$.
Fig. \ref{fig:7} shows the results obtained for
model M2k. In this case the differences are lower than 
obtained at higher frequencies for this model
(cf.\ Figs \ref{fig:3} and \ref{fig:5}).
The most surprising behaviour is present in \PULSE\ and
\LOSC. \POSC\ and \OSCROX\ also present some differences in the same
region. To understand the reason for these differences and the region
where they appear Fig. \ref{fig:8} shows the period separation $\Delta
\Pi$ between adjacent modes.  The first-order asymptotic g-mode theory
predicts a constant separation of the periods in this regime for a
given $\ell$. However, when the equations are solved numerically, this
period spacing presents several minima, and these minima are directly
linked with the mode trapping \citep{bras}.  Fig. \ref{fig:8} shows
that the position of one of these minima is the same as the position
of the largest differences.  The modes in this region have somewhat
enhanced amplitudes in the region just outside the convective core.
In \PULSE\ and \LOSC\ this apparently happens for modes somewhat
different from the remaining codes.  This is the origin of the
frequency differences.  As the mode trapping in this region for these
stellar models is related to the Brunt-V\"ais\"al\"a frequency in the
$\mu$-gradient zone, the previously mentioned inadequate treatment of
this region in model M2k is likely the reason for these
differences.

\begin{figure}
\centering
%\begin{tabular}{cc}
  \rotatebox{-90}{\includegraphics[width=6cm]{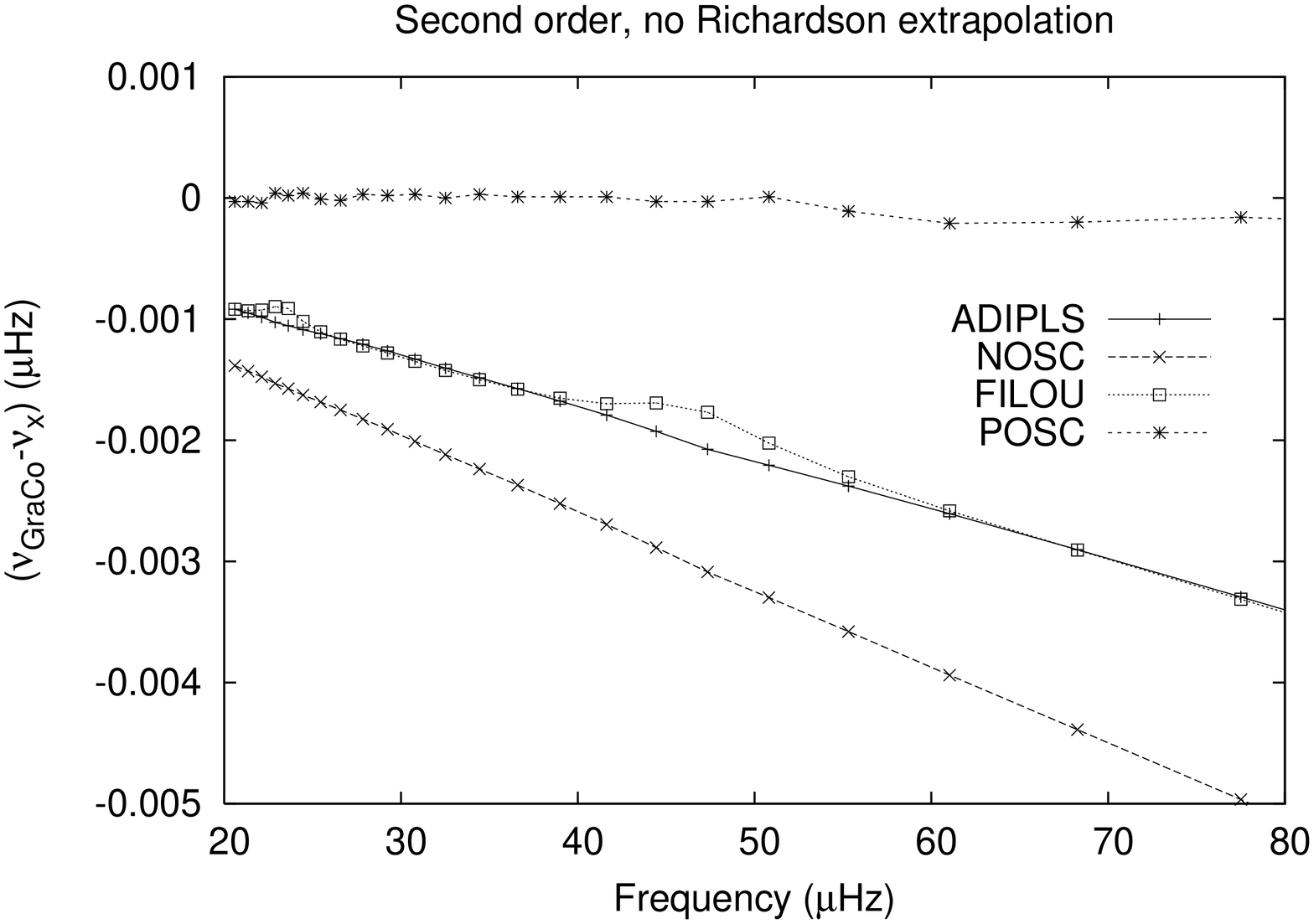}}
  \rotatebox{-90}{\includegraphics[width=6cm]{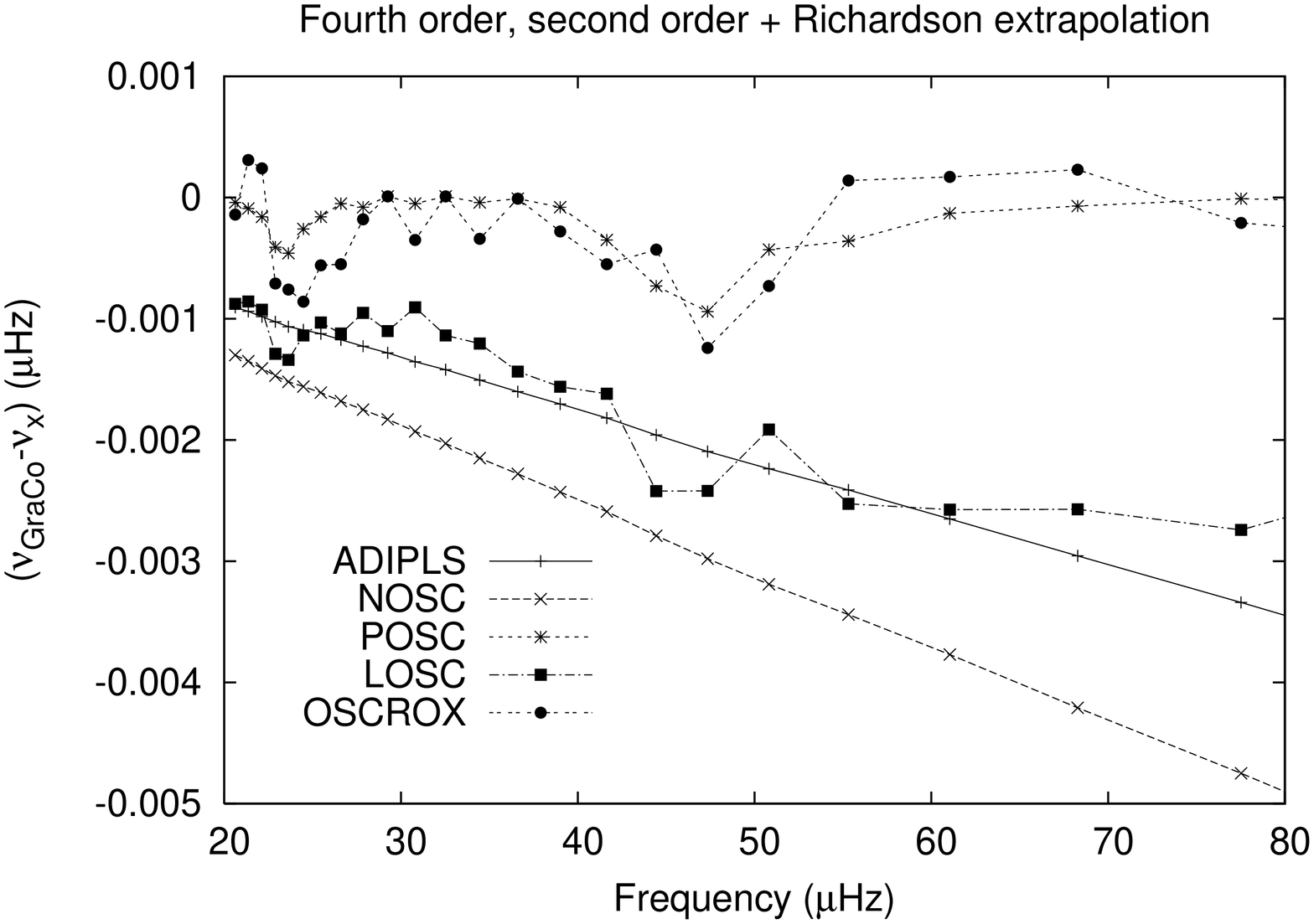}}
  \rotatebox{-90}{\includegraphics[width=6cm]{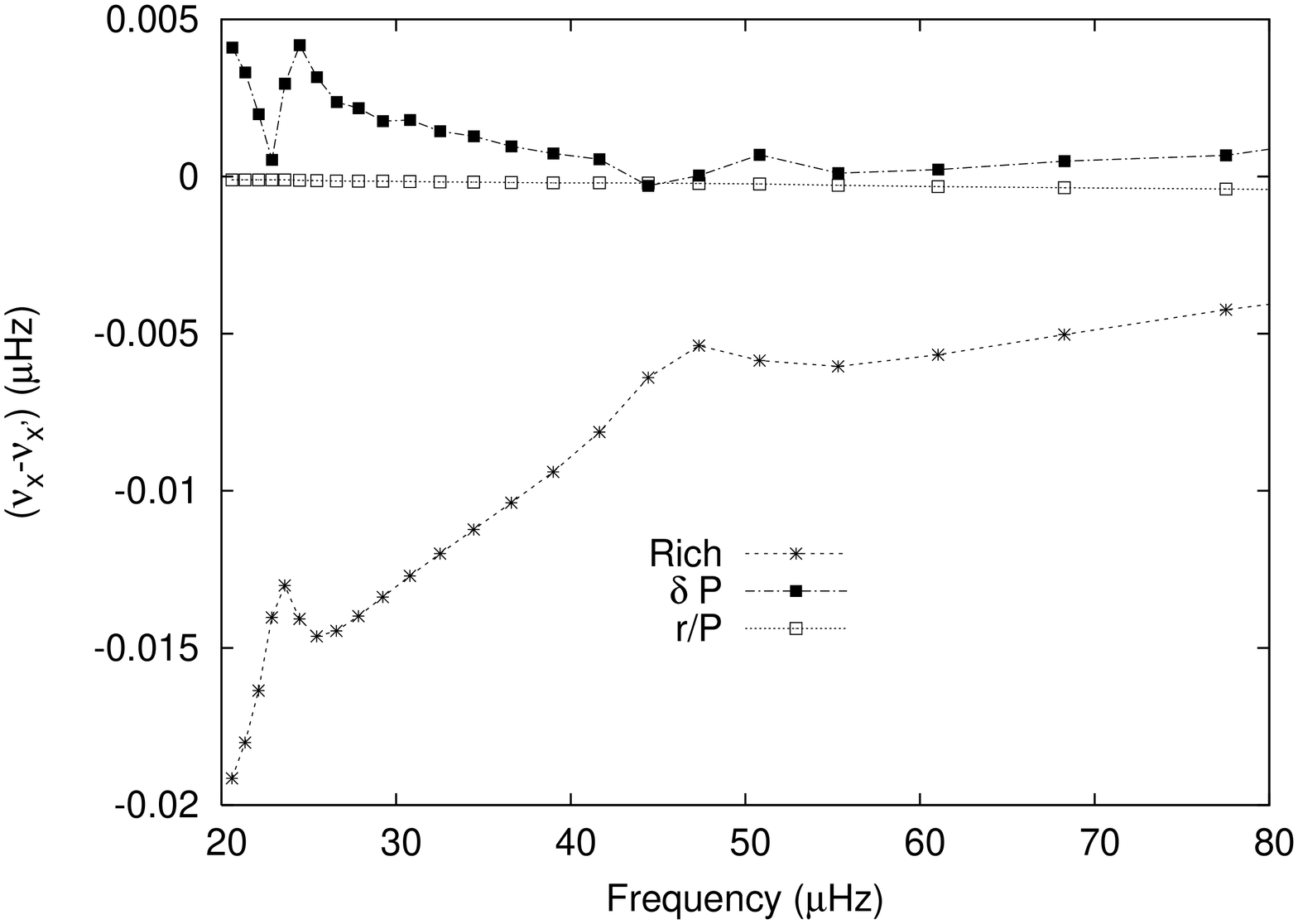}}
%\end{tabular}
\caption{Frequency comparison (reference line is \GRACO) for modes with
   $\ell=2$ as a function of the frequency obtained for
   the model M4k in the low-frequency region. In the top panel the
   models with: second order, no Richardson extrapolation, and $r$, are depicted
   (\NOSC-\ADIPLS-\GRACO-\FILOU-\POSC). The middle panel presents the
   differences obtained for the models providing fourth-order
   integration solutions or second-order plus Richardson
   extrapolation (\LOSC-\OSCROX-\NOSC-\ADIPLS-\GRACO-\POSC). In the bottom
   panel an inter-comparison for the different degrees of freedom only
   using \GRACO\ is presented.}
\label{fig:9}       % Give a unique label
\end{figure}

The top panel of Fig. \ref{fig:9} presents inter-comparisons for
$\ell=2$ frequencies calculated with a second-order scheme without
Richardson extrapolation and for model M4k,
for the same frequency range as in
Fig. \ref{fig:7}. % Here we notice that 
\POSC\ is again extremely close to \GRACO, within $2 \times 10^{-4} \, \mu$Hz.
\FILOU\ and \ADIPLS\ present very
similar results, with \NOSC\ showing slightly larger differences,
ranging from 0.0005 up to $0.0015 \,\mu$Hz relative to this group.
The differences decrease globally as far as
the frequency decreases until reaching a magnitude 
of $0.0015 \,\mu$Hz for the smallest frequency studied here.

The middle panel of Fig. \ref{fig:9} shows the same comparison for
codes using a fourth-order scheme or a second-order plus Richardson
extrapolation. Once again it can be seen that the precision resembles
the results in the previous panel. Here
\POSC\ and \OSCROX\ provide very similar results to \GRACO. \ADIPLS-\LOSC-\NOSC\
present small differences among them, with \NOSC\ having a small increasing
difference with frequency with respect to the other two.
Again the general pattern here, and in the top panel, largely reflects
the differences in $G$.
In this case we
can distinguish the codes using fourth-order integration scheme with
its apparently noisy profiles as compared with a solution using second-order
integration plus Richardson extrapolation.

\begin{figure}
\centering
%\begin{tabular}{cc}
  \rotatebox{-90}{\includegraphics[width=6cm]{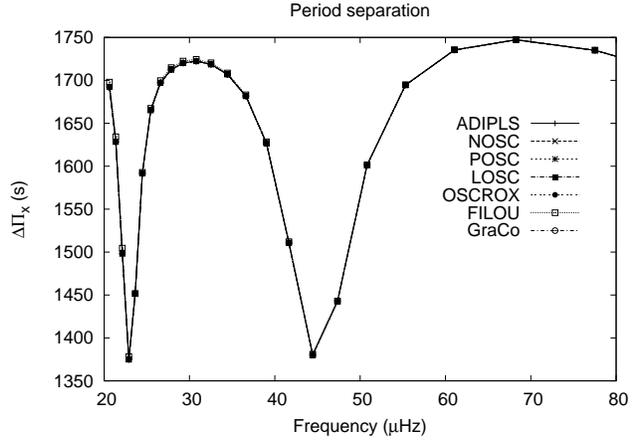}}
%\end{tabular}
\caption{Period spacing as a function of the frequency in the
  asymptotic g-mode region for the model M4k.}
\label{fig:10}       % Give a unique label
\end{figure}

In the bottom panel of Fig. \ref{fig:9}, an inter-comparison for the
different ``degrees of freedom" using only \GRACO\ is presented;
as in earlier corresponding plots the same value of $G$ is used as in
the computation of the equilibrium model.
As expected the Richardson extrapolation has a growing influence as the
frequency decreases and the mode order increases, with quite
substantial differences, compared with the differences between
different codes, for the lowest-frequency modes.  Thus, with the mesh
provided by the evolution calculation the second-order schemes have
inadequate numerical precision.  In this case,
the differences provided by the use of $r/P$ as integration
variable are negligible;
%(see Sect. ~\ref{sec:dep}).
The use of the
Lagrangian perturbation to the pressure ($\delta P$) gives rise to
frequency differences exceeding those obtained between the different
codes, at the lowest frequencies;
we note, however, that a corresponding comparison between ADIPLS and LOSC
does not show this effect which may therefore be particular to the \GRACO\
implementation.
%\note [Same comment on $\delta P$ effects as in Fig. 6, compared with the
%\ADIPLS/\LOSC\ difference].

Finally, we want to point out that in the case of model M4k, the
large differences appearing in the mode trapping region are not
found. Fig. \ref{fig:10} presents the same period separation as
Fig. \ref{fig:8} but for this model. All the codes give quite similar results.
Two mode trapping regions appear with the same frequency domain as in
Fig. \ref{fig:8}. As this model does not present any numerical
imprecision in the Brunt-V\"as\"al\"a frequency pattern, 
the obvious conclusion is that, as for the mixed modes in avoided crossing,
any numerical imprecision in the description of $N^2$ coming from the
equilibrium model can give rise to large differences in the
frequencies calculated by different oscillation codes for the g-modes
trapped in the $\mu$-gradient zone.

%-------------------------------------------------------------------
\section{Large separations (LS).}

This section is devoted to the asymptotic behaviour of p-modes through
the use of the so called ``large separations", that is, the difference
between two consecutive modes with the same spherical degree $\ell$
($\Delta=\nu(n,\ell)-\nu(n-1,\ell)$, $n$ being the radial order
of the mode). The structure of the section is similar to the previous
one. We will use the same definitions as in the previous section to study
all the frequencies ranges. From now on, the results with the M2k
model will not be discussed, since no additional information is found
from the further inter-comparisons.

%..............................................
\subsection{Large separation of radial modes.}

\begin{figure}
\centering
%\begin{tabular}{cc}
  \rotatebox{-90}{\includegraphics[width=6cm]{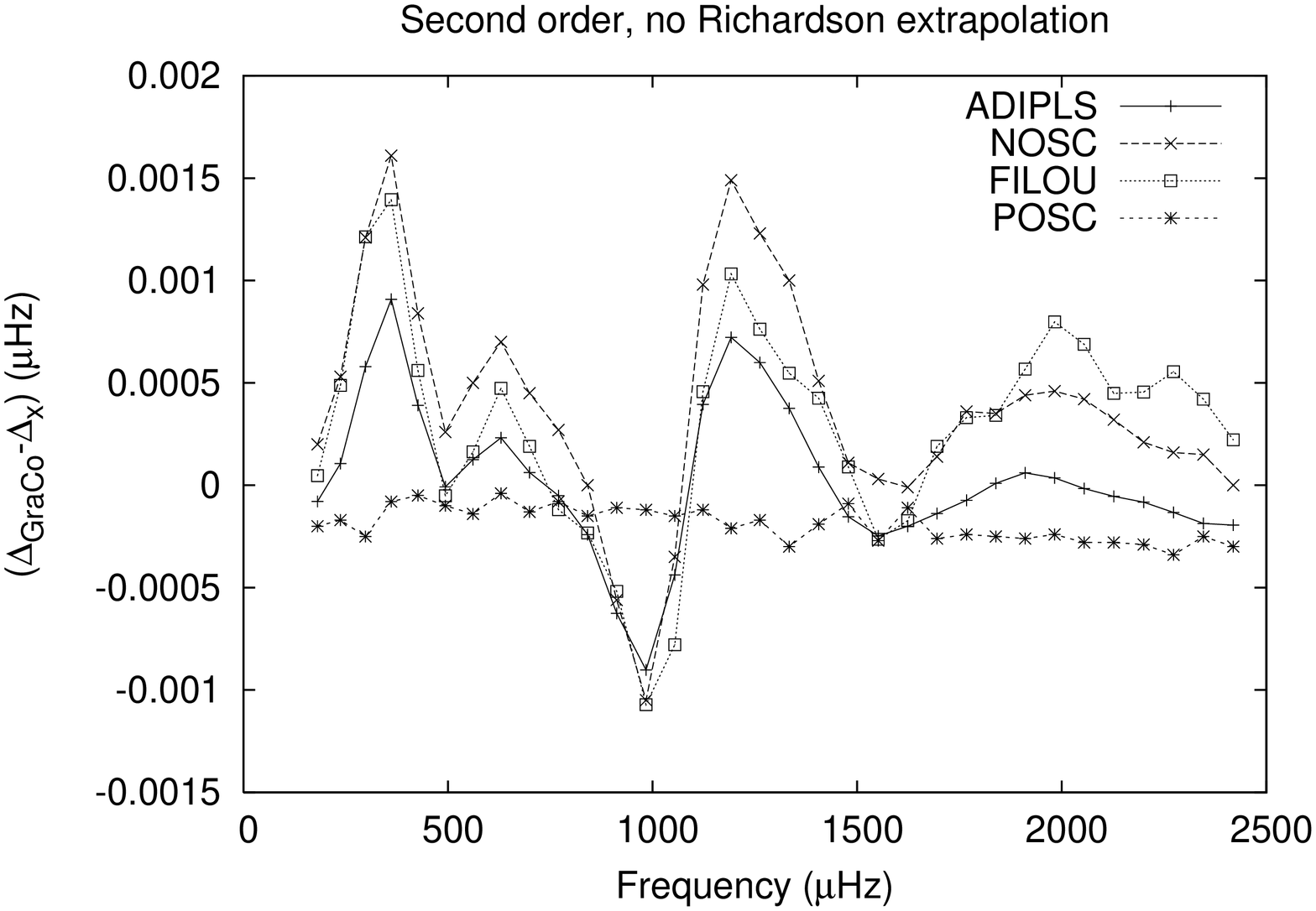}}
  \rotatebox{-90}{\includegraphics[width=6cm]{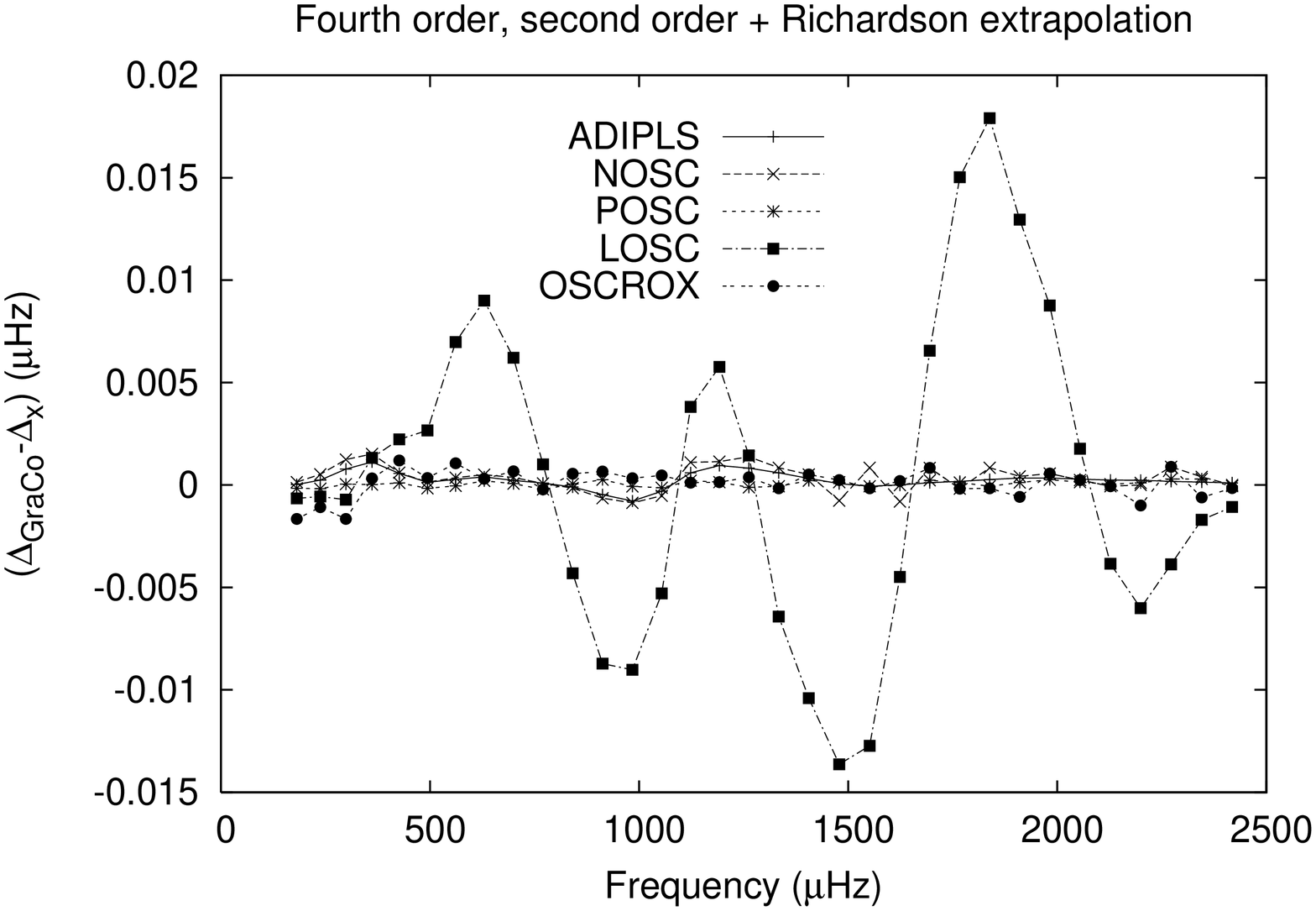}}
  \rotatebox{-90}{\includegraphics[width=6cm]{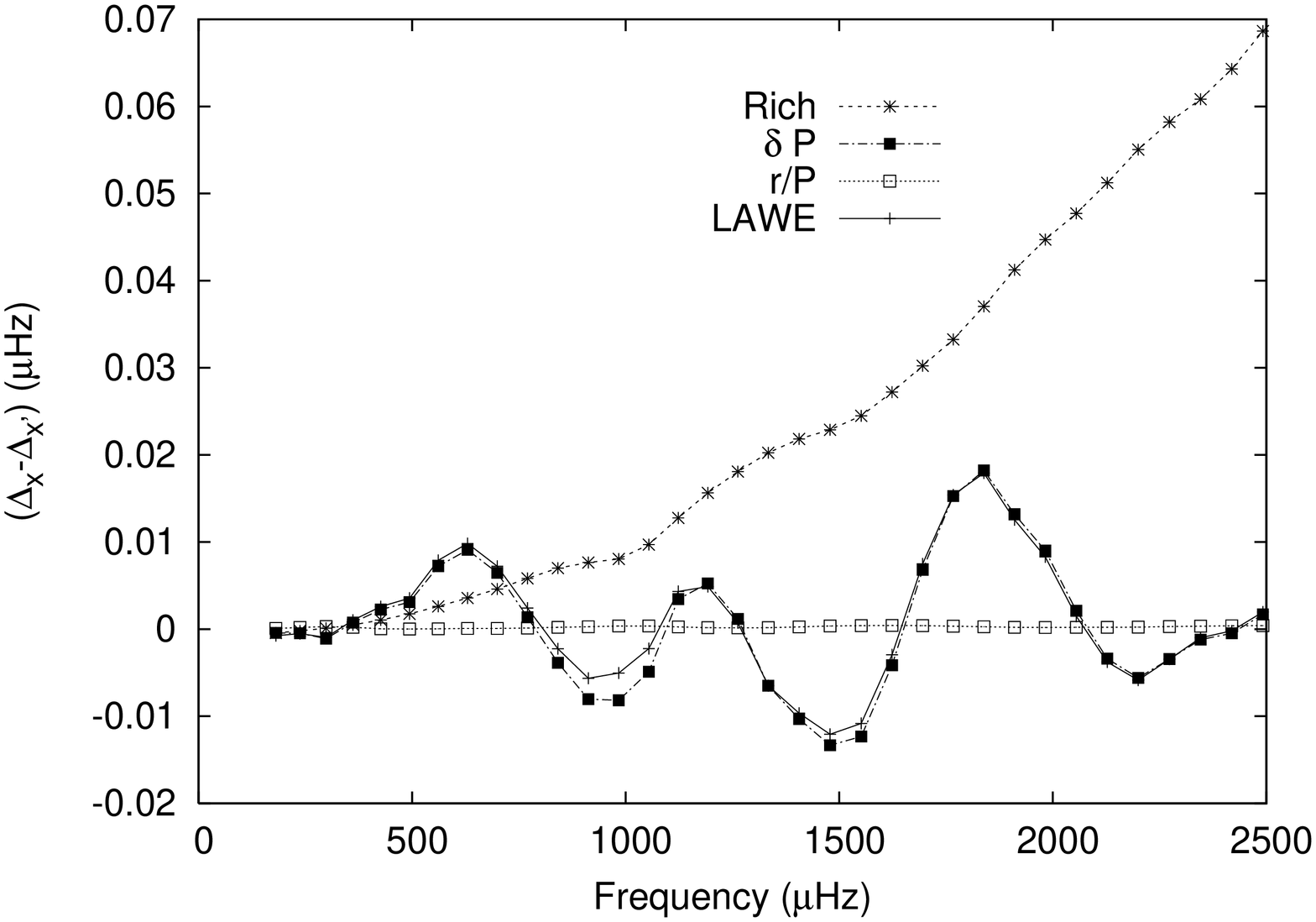}}
%\end{tabular}
\caption{Large separation inter-comparison (reference line is \GRACO)
   for modes with
   $\ell=0$ as a function of the frequency calculated
   for the model M4k. In the top panel the models with second order,
   no Richardson extrapolation, $r$,
   are depicted (\NOSC-\ADIPLS-\GRACO-\FILOU-\POSC). The
   middle panel presents the differences obtained for the models using
   fourth-order integration solutions or second-order plus
   Richardson extrapolation (\LOSC-\OSCROX-\NOSC-\ADIPLS-\GRACO-\POSC). In
   the bottom panel an inter-comparison for the different ``degrees of
   freedom" using only \GRACO\ is presented.}
\label{fig:12}       % Give a unique label
\end{figure}

Figure \ref{fig:12} shows the results obtained for model M4k. In the top
panel differences for the different codes using second order, no
Richardson extrapolation, $r$, are depicted.
\POSC\ and \GRACO\ are again extremely close.
The rest of the codes
(\ADIPLS-\NOSC-\FILOU-\GRACO) present differences of
up to around $0.0015 \,\mu$Hz, generally sharing an oscillatory pattern,
particularly at relatively low frequency;
we have no explanation for this behaviour.
However, the effect is evidently small.

In the middle panel of Fig. \ref{fig:12} the LS differences for the
different codes, using a fourth-order integration scheme or a
second-order plus Richardson extrapolation, are presented. With the
exception of \LOSC\ the global
behaviour is similar to that obtained without Richardson
extrapolation,
the \NOSC-\ADIPLS-\OSCROX-\GRACO\ differences being always 
lower than $0.002 \,\mu$Hz.
The pattern of the \LOSC\ differences,
presenting differences one order of
magnitude larger than the rest of the codes,
is clearly related to the corresponding oscillatory pattern
found in Fig.~\ref{fig:4};
as before it is explained by the use of the LAWE differential equation.

The bottom panel of Fig. \ref{fig:12} shows the LS differences for the
radial modes obtained with \GRACO\ and model M4k when different ``degrees
of freedom" are used.
The Richardson extrapolation introduces
differences increasing with frequency and mode order,
as expected, giving the largest
differences, around $0.07 \,\mu$Hz.
This again emphasizes the inadequacy of the second-order schemes for
the highest-order modes, on the M4k mesh.
The integration variable $r/P$ gives
differences slightly lower than $5\times 10^{-4} \,\mu$Hz, i.e.,
much smaller than that found for different codes using the same
numerical techniques, as depicted in the previous panels.
The use of the Lagrangian perturbation to the pressure $\delta P$ 
and the LAWE differential equation
show the same oscillating behaviour and values as those previously
observed for the \LOSC\ results.
As discussed above, this is related to the inconsistency in $A^*$ in 
model M4k (see Sect.~\ref{sec:dep}).

%..............................................
\subsection{Non-radial modes with $\ell=2$}

To illustrate the differences appearing in the case of non-radial
modes, the spherical degree $\ell=2$ has been arbitrarily chosen. We
have divided the frequency spectrum in three regions, like in the
direct frequency inter-comparison: 1) high-frequency region, 2)
frequencies around the fundamental radial mode and 3) low-frequency
region. In all cases, a study similar to that developed in the radial
case has been carried out. In the low-frequency region, the more
physical period separation is studied, instead of the
frequency separation relevant for acoustic modes.

In the first region the results are very similar to those obtained for
the radial case; therefore the plots are not presented here. As in the
direct frequency inter-comparison case, \LOSC\ also presents an
oscillating pattern, owing to the use of $\delta P$ as eigenfunction
(see Sect.~\ref{sec:dep}).  On the other hand, the only noticeable
difference, when compared with the radial case in this region, is that
for $\ell=2$ the precision among codes using the same integration
procedures is slightly higher.

\begin{figure}
\centering
%\begin{tabular}{cc}
  \rotatebox{-90}{\includegraphics[width=6cm]{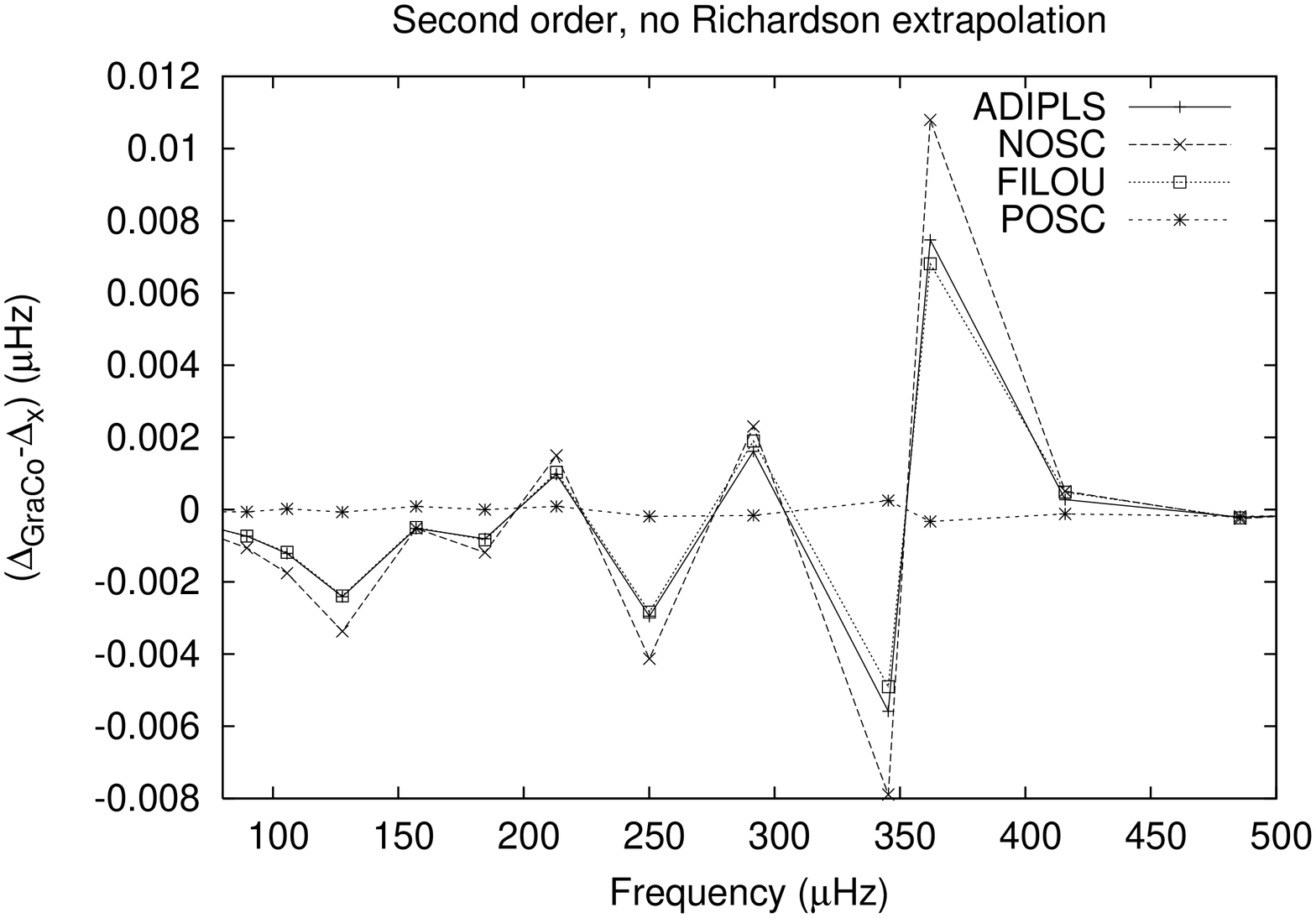}}
  \rotatebox{-90}{\includegraphics[width=6cm]{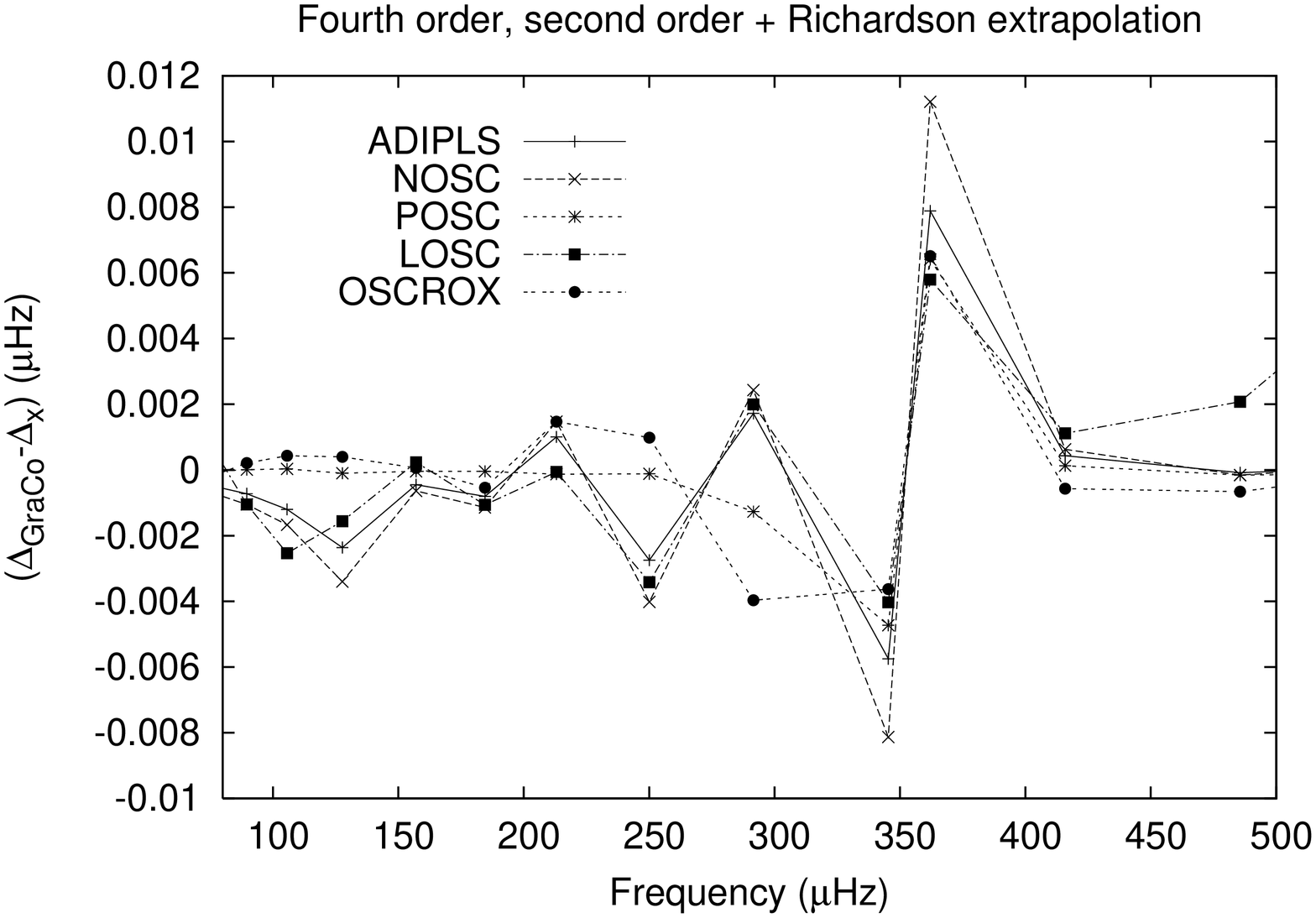}}
  \rotatebox{-90}{\includegraphics[width=6cm]{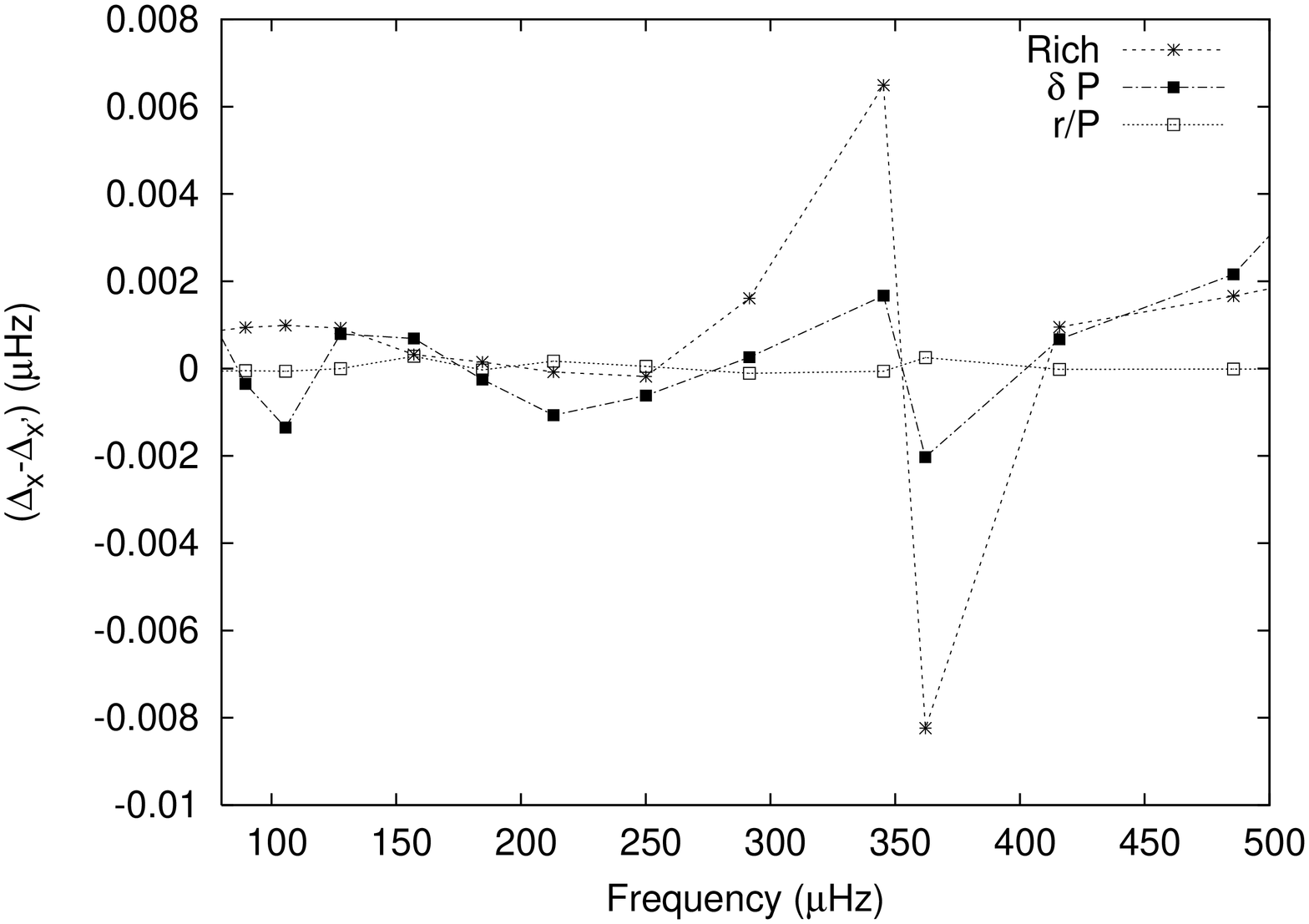}}
%\end{tabular}
\caption{Large separation inter-comparison (reference line is \GRACO)
   for modes with
   $\ell=2$ as a function of the frequency calculated
   for the model M4k and for modes around the fundamental radial
   one. In the top panel the models with second order, 
   no Richardson extrapolation,
   $r$, are depicted (\NOSC-\ADIPLS-\GRACO-\FILOU-\POSC). The middle panel
   presents the differences obtained for the models using fourth-order
   integration solutions or second-order plus Richardson
   extrapolation (\LOSC-\OSCROX-\NOSC-\ADIPLS-\GRACO-\POSC). In the bottom
   panel an inter-comparison for different ``degrees of freedom" using
   only \GRACO\ is presented.}
\label{fig:14}       % Give a unique label
\end{figure}

The results obtained with the codes using a second-order scheme,
%in the case of the model M4k,
for the modes around the fundamental radial
mode, are depicted in the top panel of Fig. \ref{fig:14}.
\POSC\ remains very close to \GRACO,
with the other set of codes (\FILOU-\NOSC-\ADIPLS) extending the
oscillatory pattern in the top panel of Fig.~\ref{fig:12},
with the largest difference being $0.01 \,\mu$Hz for the mixed modes.
This set of codes agrees to within differences around $0.001 \,\mu$Hz
(slightly larger for \NOSC).

The middle panel of Fig. \ref{fig:14} depicts the differences for a
fourth-order integration scheme or second-order plus Richardson extrapolation.
The precision of the different codes is similar to that given by the
second-order scheme without Richardson extrapolation.
The maximum difference is
lower than $0.08 \,\mu$Hz, and most of the codes present differences
around $0.005 \,\mu$Hz. Once again the largest differences are obtained
in the mixed modes. No different behaviours depending on
the integration scheme are found.

\begin{figure}
\centering
%\begin{tabular}{cc}
  \rotatebox{-90}{\includegraphics[width=6cm]{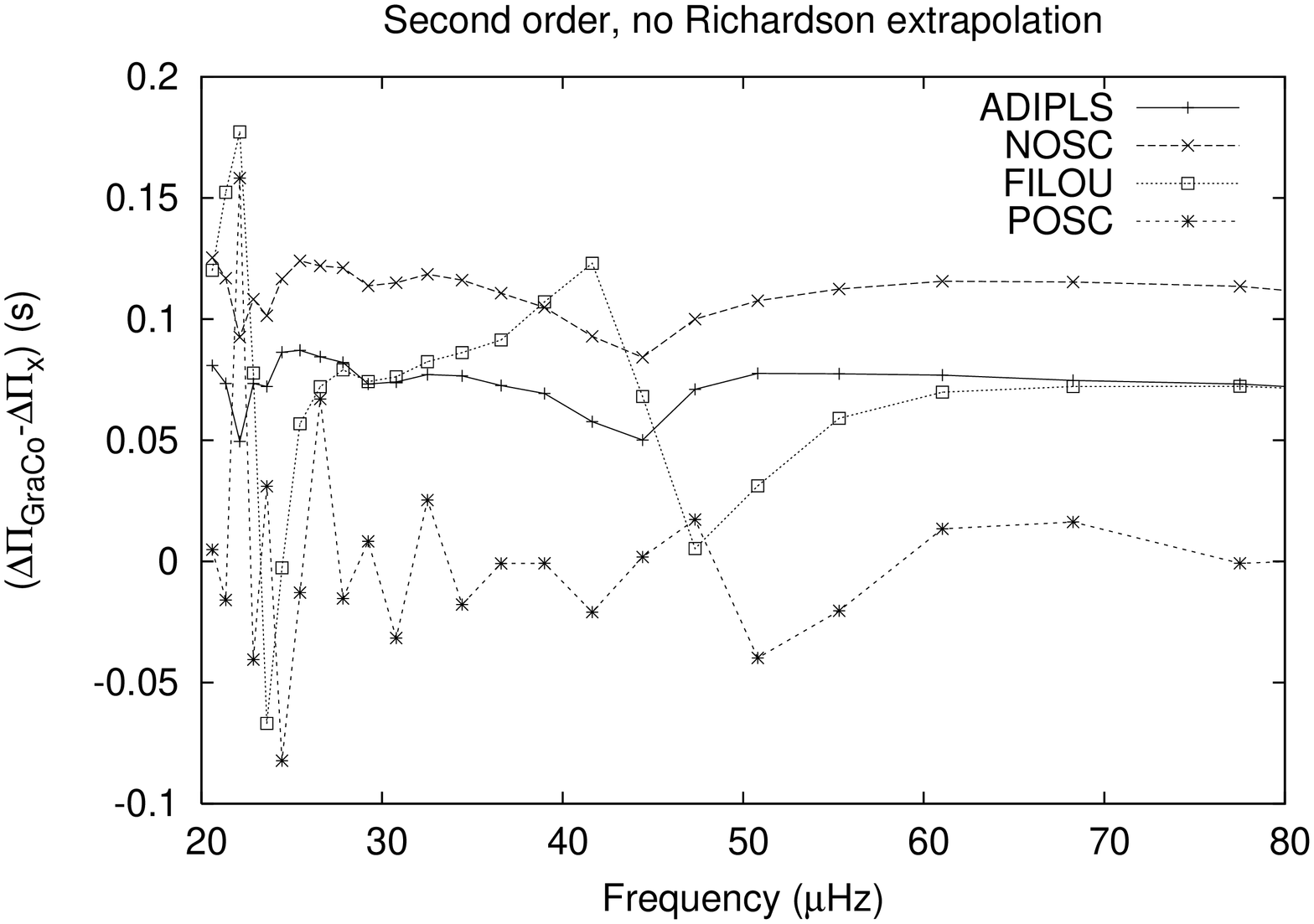}}
  \rotatebox{-90}{\includegraphics[width=6cm]{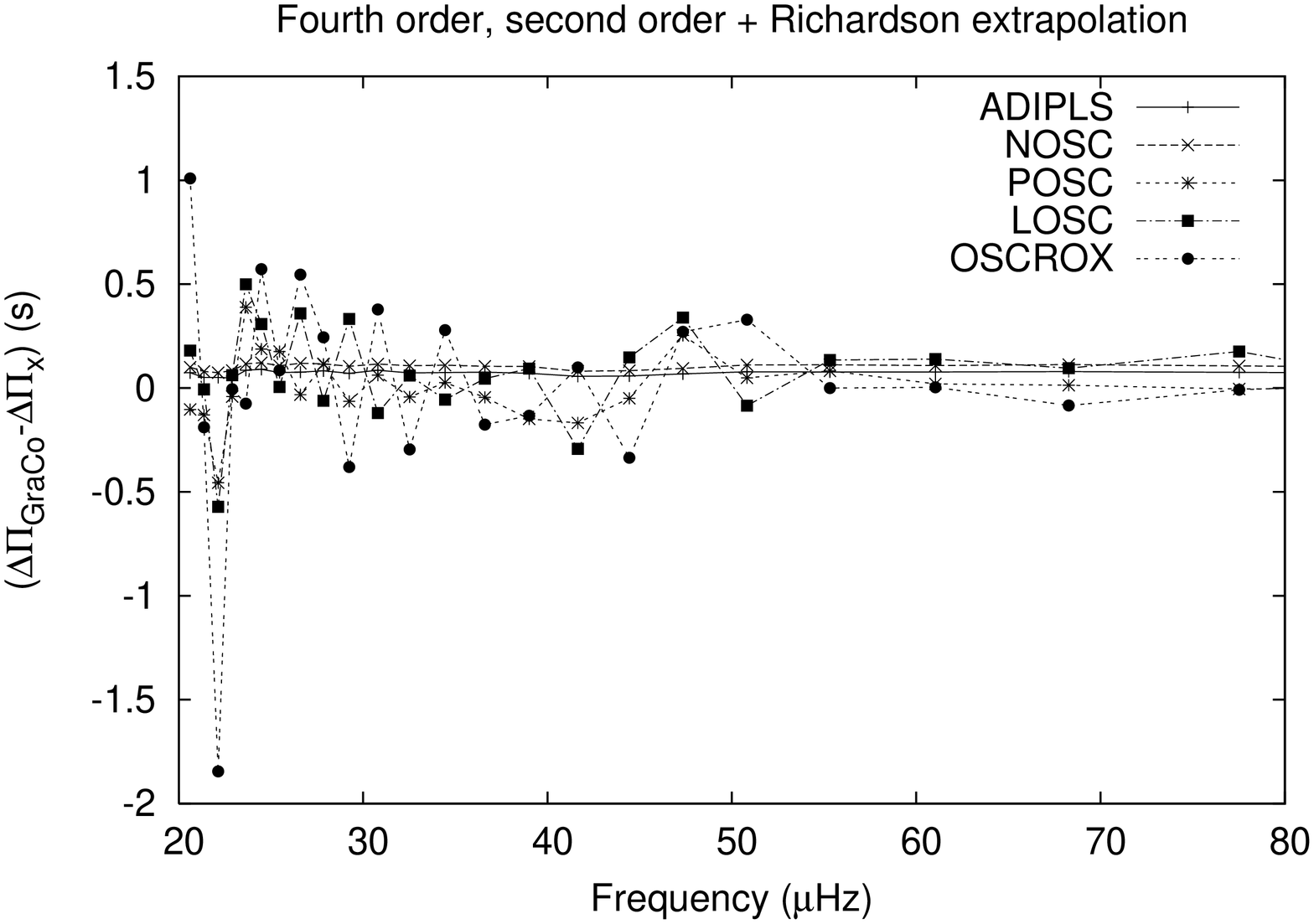}}
  \rotatebox{-90}{\includegraphics[width=6cm]{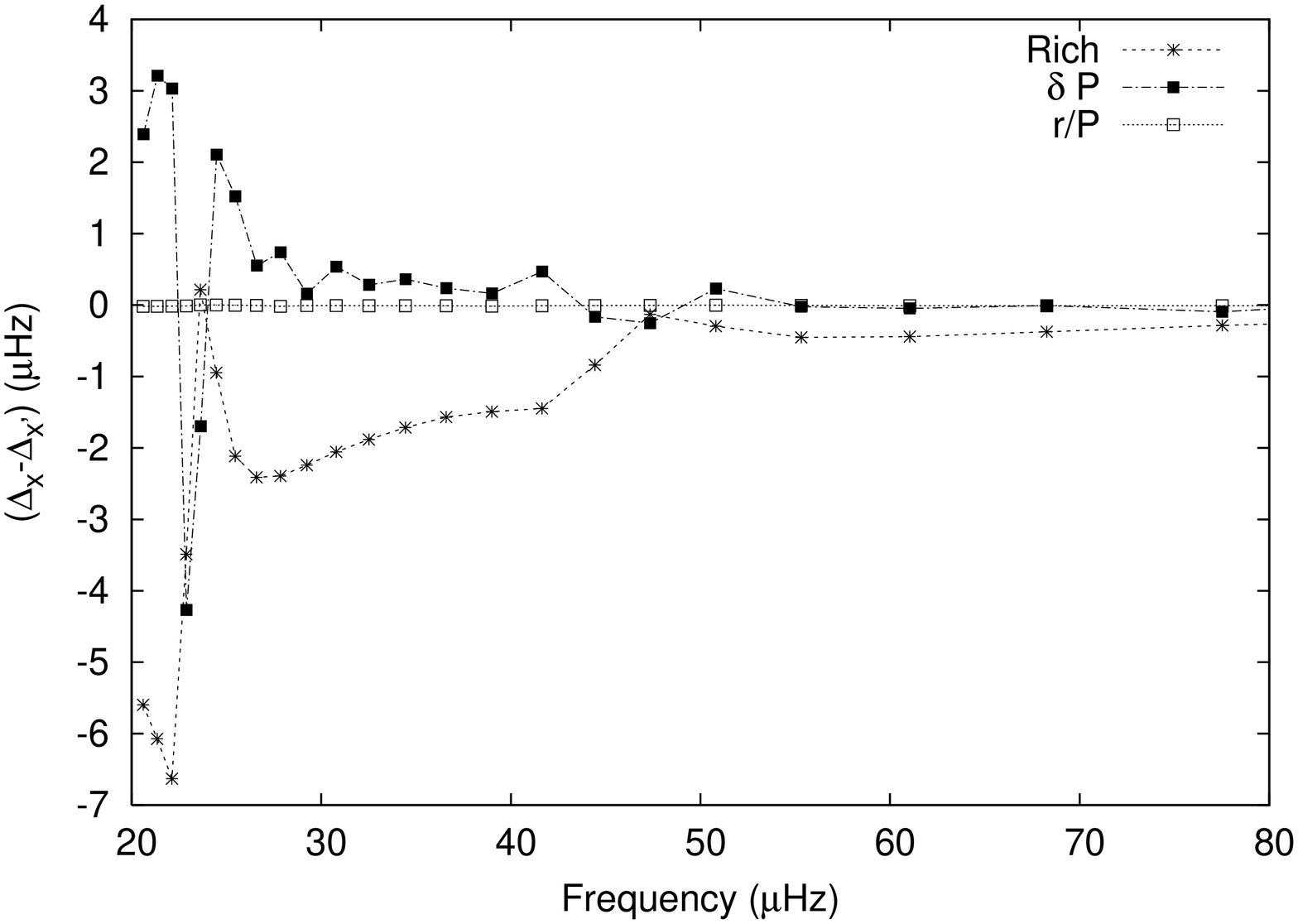}}
%\end{tabular}
\caption{Period separation inter-comparison (reference line is \GRACO)
   for modes with
   $\ell=2$ as a function of the frequency calculated
   for the model M4k 
   in the low-frequency region. In the top panel the
   models with second order, no Richardson extrapolation, $r$, are depicted
   (\NOSC-\ADIPLS-\GRACO-\FILOU-\POSC). The middle panel presents the
   differences obtained for models using fourth-order integration
   solutions or second-order plus Richardson extrapolation
   (\LOSC-\OSCROX-\NOSC-\ADIPLS-\GRACO-\POSC). In the bottom panel an
   inter-comparison for the different choices of ``degrees of freedom"
   using only \GRACO\ is presented.}
\label{fig:16}       % Give a unique label
\end{figure}

The bottom panel of Fig. \ref{fig:14} is devoted to the differences
obtained with \GRACO\ when different options for the solution of the
differential equations are chosen. In this frequency region the replacement
of $r$ by $r/P$ as integration variable introduces the smallest
differences, one order of magnitude lower than those obtained
using the codes with the same ``degree of freedom". 
Using the Lagrangian
perturbation to the pressure ($\delta P$) as eigenfunction
causes some difference indicating sensitivity to whether or not $A^*$ is used.
The Richardson extrapolation
introduces differences lower than those found for the different codes,
except in the avoided crossing zone of mixed modes, 
where changes comparable to the largest one are found.
Given the rather substantial variations in
the bottom panel, and the fact that most codes show the same variation in the 
top two panels, one might suspect that the dominant source of this variation
is in fact in the \GRACO\ results used as reference.

The low-frequency region is studied through the period separation for
$\ell = 2$ ($\Delta \Pi$ in seconds), illustrated in
Fig. \ref{fig:16}. The top panel of this figure shows the differences
found using only codes with a second-order integration
scheme. \ADIPLS\ and \NOSC\ show similar shifts of around 0.1~s
relative to \GRACO.  \FILOU\ is similar at the higher frequencies but
shows a small oscillating behaviour in the mode-trapping regions.
Finally \POSC\ presents a quite noisy pattern, varying around zero.
Again the overall grouping of the differences (\ADIPLS-\NOSC-\FILOU\
and \GRACO-\POSC) reflects the different values of $G$.  

Results obtained using a fourth-order integration scheme or a
second-order plus Richardson extrapolation are compared in the middle
panel of Fig. \ref{fig:16}. The values of the differences found in
this case are of the same order as in the previous
inter-comparison. Here we can distinguish codes using a fourth-order
scheme or a second-order plus Richardson extrapolation, because of the
apparently random pattern they present, with differences one order of
magnitude larger than the main values. The frequency differences
obtained with the \OSCROX\ and \LOSC\ results are those presenting a
noisy behaviour, when comparing with a second-order plus Richardson
extrapolation solution (as \GRACO\ does) as the reference line.
\POSC\ also presents some differences in the mode-trapped regions when
compared with other codes but using the same integration scheme.

The bottom panel of Fig. \ref{fig:16} shows the differences obtained
with the same code (\GRACO) and different choices of the ``degrees of
freedom".  The use of the Richardson extrapolation introduces
substantial differences, of the order of seconds, with noticeable
wiggles in the two mode-trapping regions.  As noted previously this
reflects the inadequacy of the second-order schemes for high-order
modes.  The use of the Lagrangian perturbation to the pressure as
variable ($\delta P$) also introduces substantial differences,
particularly around the trapped modes near $22 \, \mu$Hz,
related to the frequency differences found with \GRACO\ in this region
when $\delta P$ is used (cf.\ Fig.~\ref{fig:9}).
Using $r/P$ as
integration variable gives a small difference when compared with $r$,
one order of magnitude lower than the differences found
between the codes using the same numerical integration schemes.

%-------------------------------------------------------------------
\section{Small separations (SS).}

In this section we study the inter-comparisons for the small
separations (SS) 
$\delta_{\ell \ell^\prime} \equiv\nu(n, \ell)-\nu(n{-}1, \ell{+}2)$.
Therefore, two sets of inter-comparisons can be done, one for the
\hbox{($\ell=0-\ell=2$)} modes and another for the ($\ell=1-\ell=3$)
modes. In this case only the high-frequency region is studied,
since this is where this quantity has physical meaning.
As in the previous section we concentrate on results for model M4k.

%..............................................
\subsection{Small separations $\delta_{02}$}

\begin{figure}
\centering
%\begin{tabular}{cc}
  \rotatebox{-90}{\includegraphics[width=6cm]{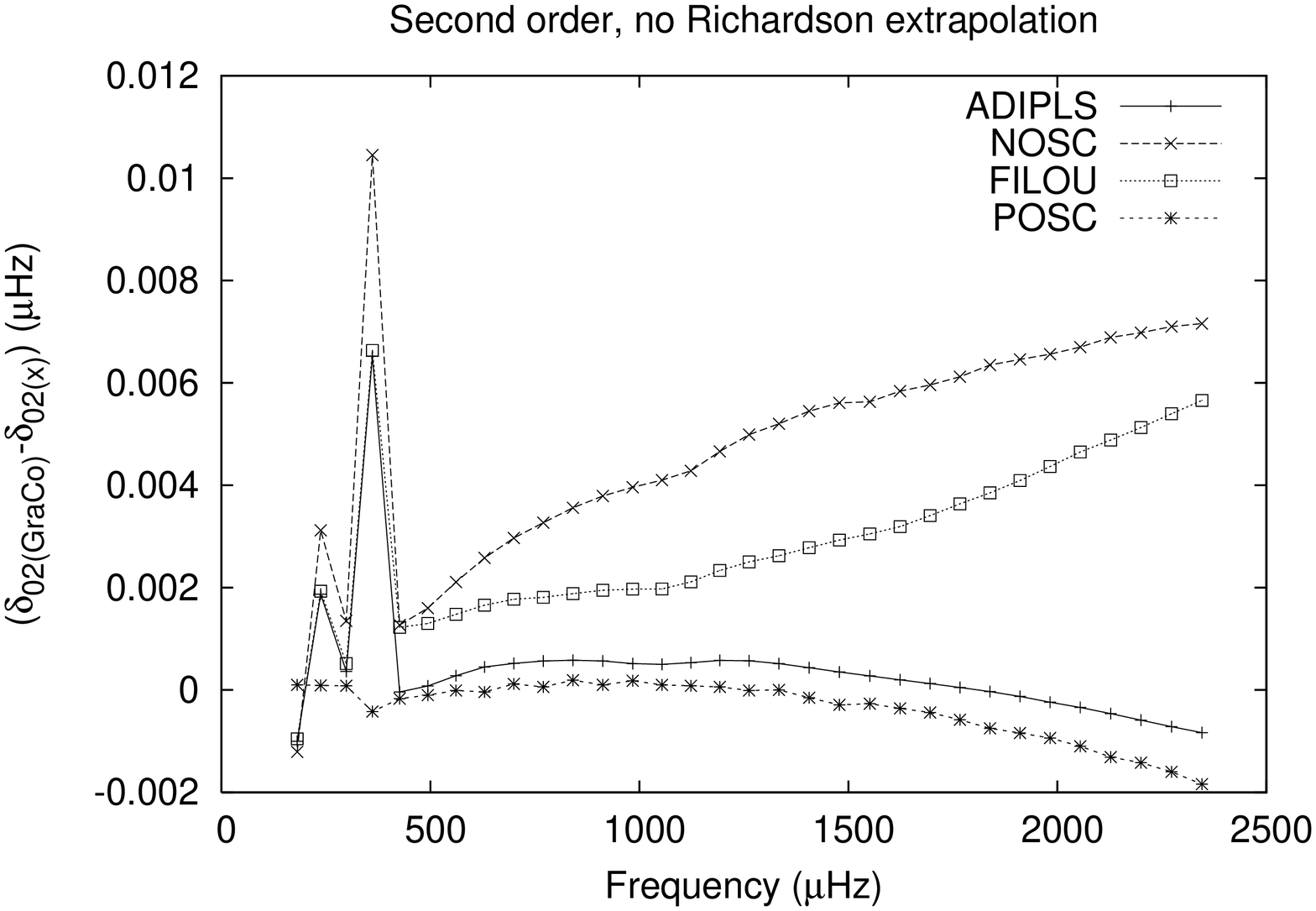}}
  \rotatebox{-90}{\includegraphics[width=6cm]{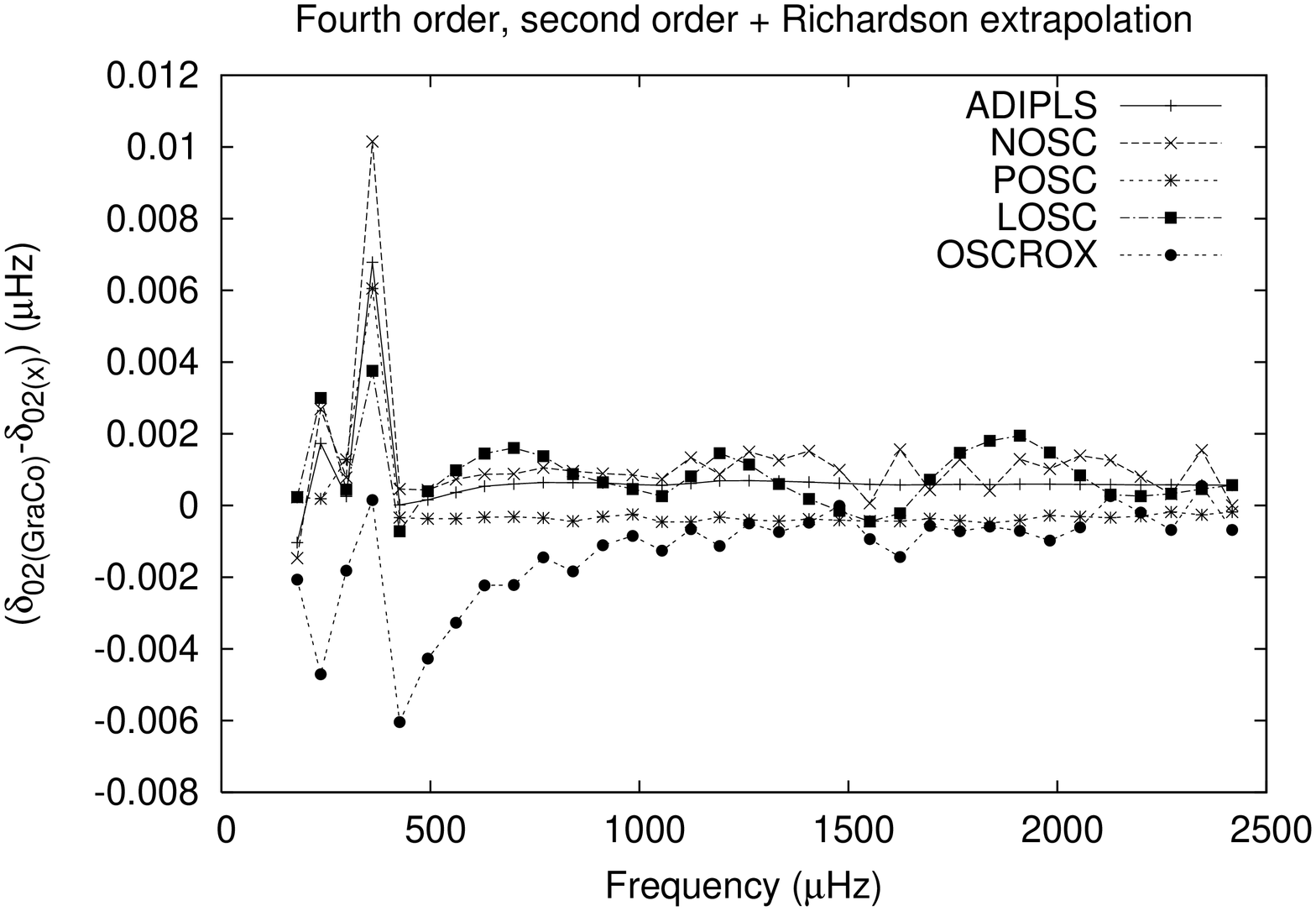}}
  \rotatebox{-90}{\includegraphics[width=6cm]{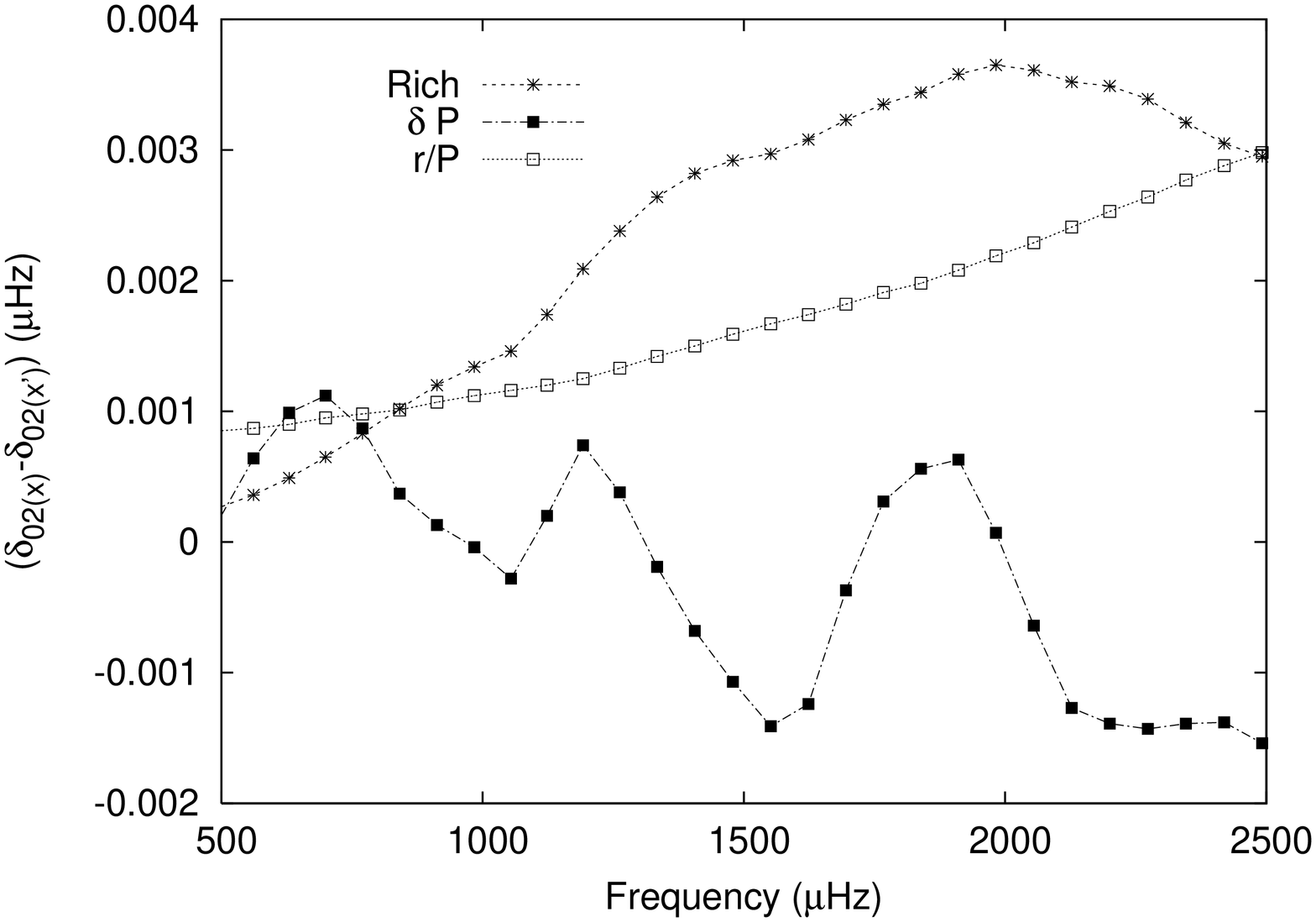}}
%\end{tabular}
\caption{Small separation inter-comparison (reference line is \GRACO)
   for modes with
   $\ell=0-2$ as a function of the frequency calculated
   for the model M4k in the high-frequency region. In the top panel models
   with second order, no Richardson extrapolation, $r$, are depicted
   (\NOSC-\ADIPLS-\GRACO-\FILOU-\POSC). The middle panel presents the
   differences obtained for models using fourth-order integration
   schemes or second-order plus Richardson extrapolation
   (\LOSC-\OSCROX-\NOSC-\ADIPLS-\GRACO-\POSC). In the bottom panel an
   inter-comparison for different ``degrees of freedom" only using
   \GRACO\ is presented.}
\label{fig:18}       % Give a unique label
\end{figure}

The results obtained for the small separation $\delta_{02}$
%with the model Mk4
are presented in Fig. \ref{fig:18}. The top panel of this
figure shows the differences obtained with the codes solving the set
of differential equations with a second-order integration scheme. In
the avoided crossing region some wiggles occur related to variations
for $\ell=2$, but these wiggles are of the order of magnitude of the
differences obtained for the high-frequency region. \ADIPLS-\POSC-\GRACO\
present similar values for high frequencies and \NOSC-\FILOU\ presents
differences around $0.004 \,\mu$Hz, and always lower than $0.01
\,\mu$Hz. These differences are much lower than the expected
observational errors for CoRoT.

The middle panel shows the same inter-comparison using a fourth-order
integration scheme or a second-order plus Richardson extrapolation.
Once again we cannot distinguish the integration scheme used. Just
\LOSC\ shows an oscillatory pattern for high frequencies due to the use
of LAWE or $\delta P$ (see Sect.~\ref{sec:dep}).  The order of
magnitude of the main differences is the same as those obtained using
only a second-order scheme. The wiggles in the avoided crossing
regions are still present, and for high frequencies the differences
are all in the range $[-0.001,0.001] \,\mu$Hz. \ADIPLS\ and \POSC\ present
an almost constant difference with \GRACO, \NOSC\ and \OSCROX\ show a
small noisy behaviour.

The bottom panel of Fig. \ref{fig:18} presents the SS differences
induced in \GRACO\ when different choices of ``degrees of freedom" are
selected.
In this case they are all below, or of the same order of magnitude as,
the spread between the different codes illustrated in the middle panel.
Interestingly, Richardson extrapolation introduces differences
for higher frequencies far smaller than found for the large separation
(Fig.~\ref{fig:12}).
%\note [Note the change: the previous statement must have referred to
%earlier results].
The use of the Lagrangian perturbation to the pressure ($\delta P$) as
variable results in the same oscillatory pattern as seen for \LOSC\ in
the middle panel.
%For those around the fundamental radial mode the variations
%are even larger than those found using the Richardson extrapolation.
The integration variable $r/P$ gives increasing differences in the
range [0.001,0.003] $\mu$Hz,
%a large wiggle for the avoided
%crossing modes, and constant differences for the rest of the spectrum,
similar to the Richardson extrapolation,
probably reflecting a difference in the sensitivity of radial and nonradial
modes to the choice of independent variable in the \GRACO\ code;
however, the effect is obviously small.
Note that in this case the use of the LAWE for the radial modes, keeping the
default ``degrees of freedom" (including the use of $P'$) for $\ell = 2$,
is hardly meaningful; thus no results are included for LAWE.
%These differences are close to the expected
%observational accuracies (for example, $0.1 \,\mu$Hz for COROT).

%..............................................
\subsection{Small separations $\delta_{13}$}

\begin{figure}
\centering
%\begin{tabular}{cc}
  \rotatebox{-90}{\includegraphics[width=6cm]{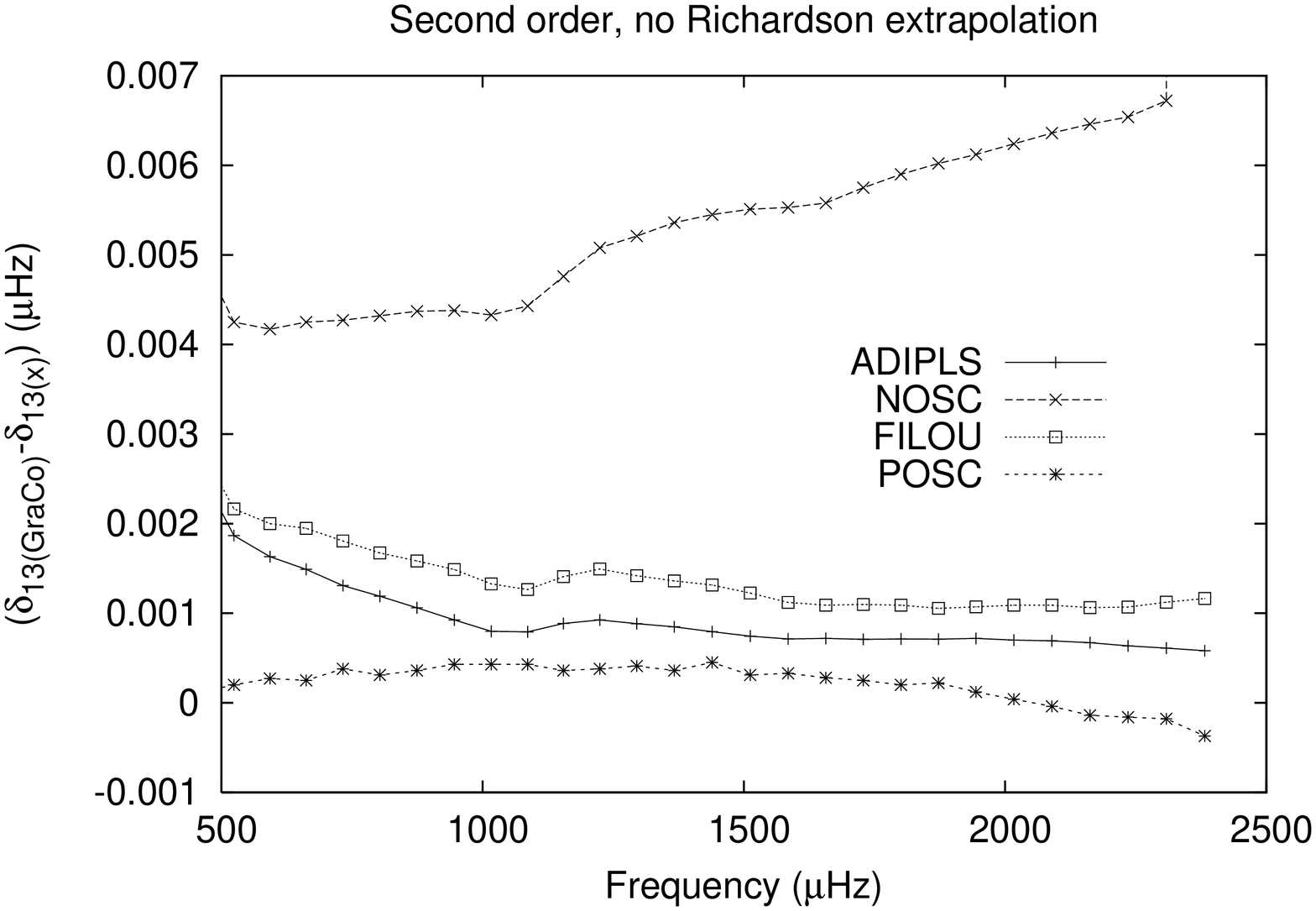}}
  \rotatebox{-90}{\includegraphics[width=6cm]{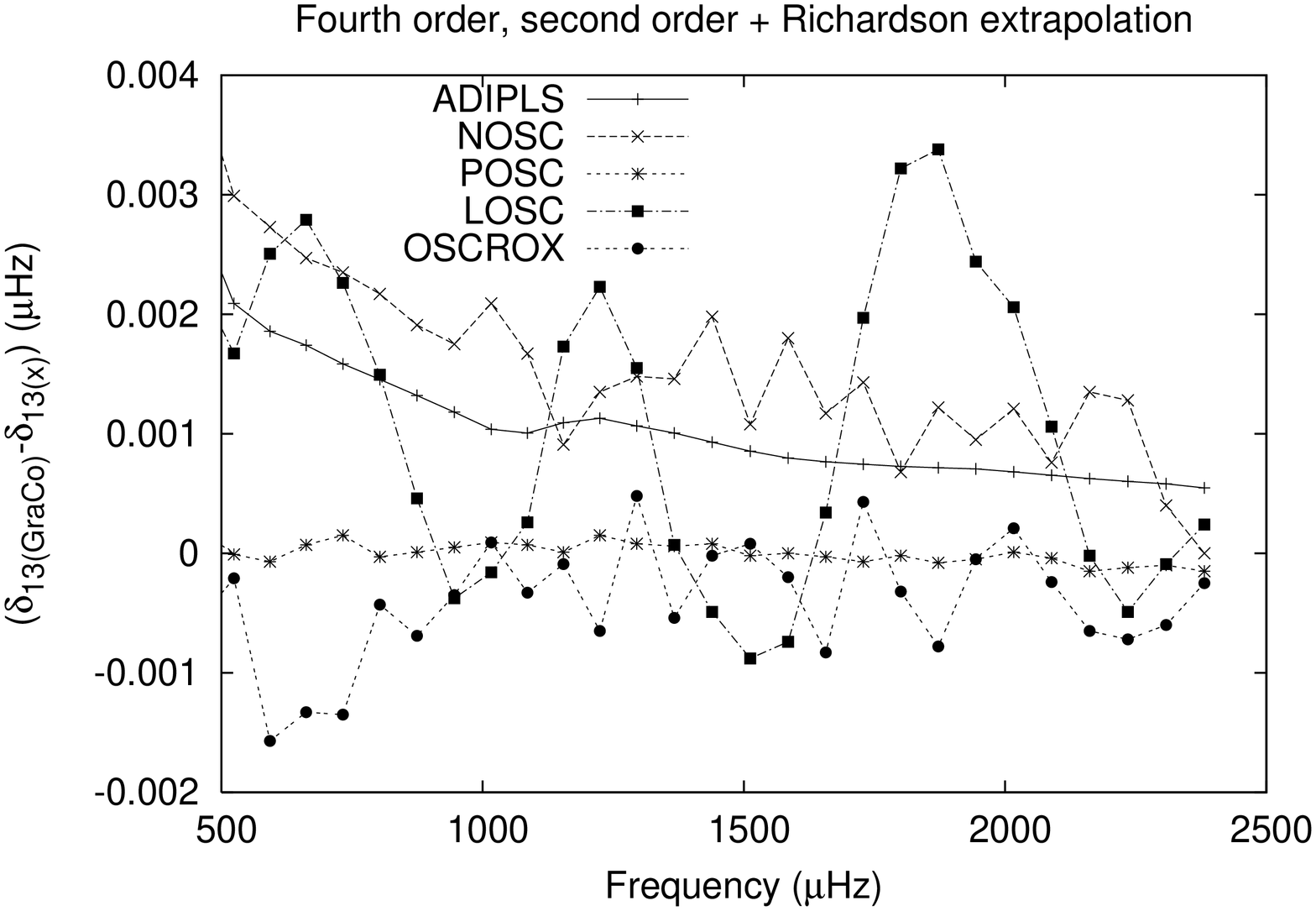}}
  \rotatebox{-90}{\includegraphics[width=6cm]{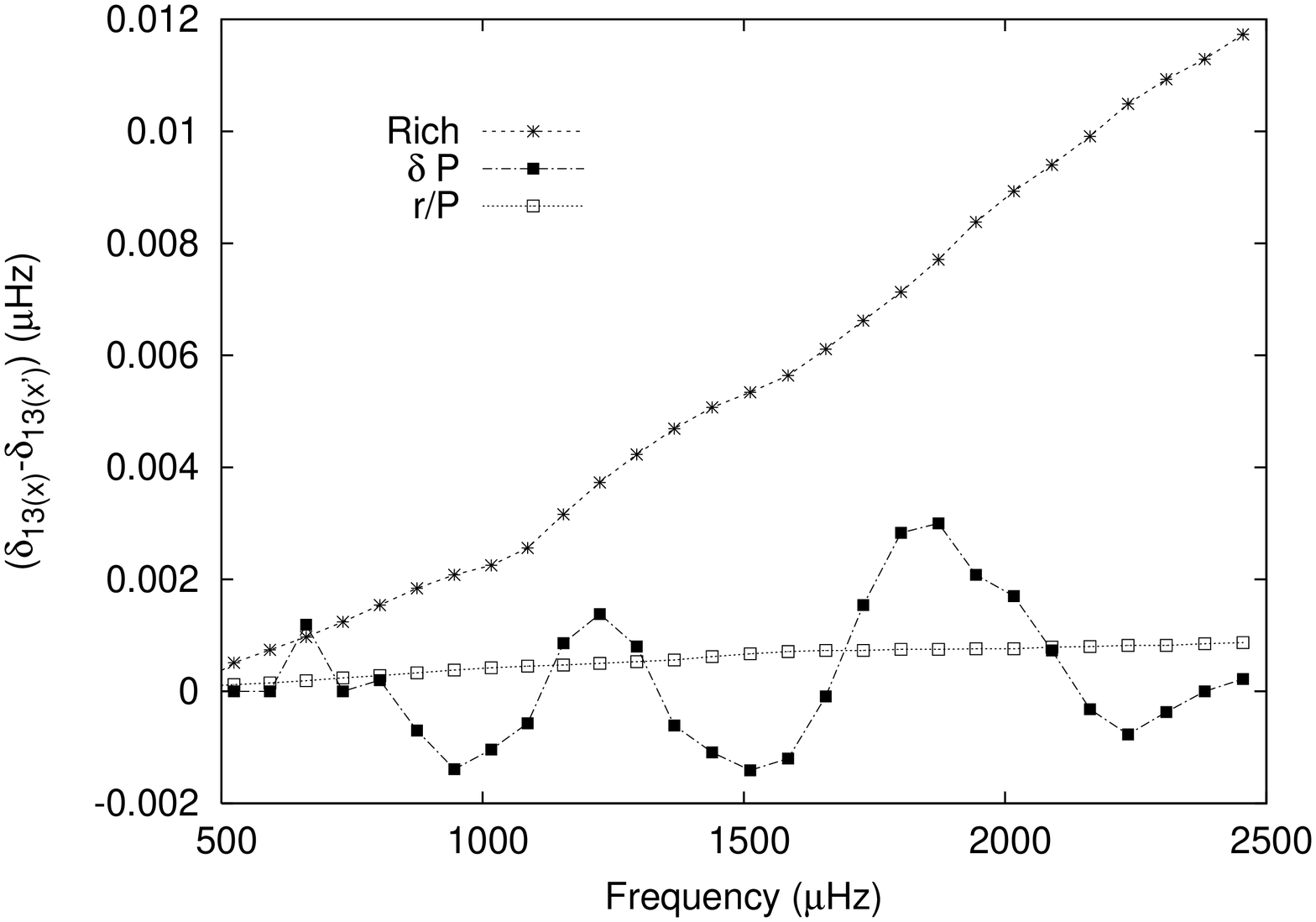}}
%\end{tabular}
\caption{SS inter-comparison (reference line is \GRACO) for modes with
   $\ell=1-3$ as a function of the frequency obtained
   for model M4k in the high-frequency region. In the top panel
   the models with second order, no Richardson extrapolation, $r$, are depicted
   (\NOSC-\ADIPLS-\GRACO-\FILOU-\POSC). The middle panel presents the
   differences calculated for models using fourth-order integration
   solutions or second-order plus Richardson extrapolation
   (\LOSC-\OSCROX-\NOSC-\ADIPLS-\GRACO-\POSC). In the bottom panel an
   inter-comparison of the different ``degrees of freedom" using only
   \GRACO\ is presented.}
\label{fig:20}       % Give a unique label
\end{figure}

In Fig. \ref{fig:20} the inter-comparison of $\delta_{13}$ is done for
model M4k. The top panel shows the differences given by the codes
using a second-order integration scheme. For \POSC-\ADIPLS-\FILOU-\GRACO\
the differences are between $-0.001$ and $0.002 \,\mu$Hz
and \NOSC\ shows
somewhat larger values. In any case the precision is good and the
patterns rather smooth.

The codes using a fourth-order integration scheme or a second-order
plus Richardson extrapolation have produced the results presented in the middle
panel of this figure. The precision is very similar as in the previous
panel or even a little higher. \LOSC\ shows the oscillatory pattern
already obtained in all the previous inter-comparisons. \POSC\ and
\ADIPLS\ present smooth difference profiles when compared with \GRACO. On
the other hand \OSCROX\ and \NOSC\ give a low noisy profile. Once again the
integration scheme used cannot be discriminated.

Finally, the bottom panel presents the effect in $\delta_{13}$ of
using different numerical integration schemes.  Using Richardson
extrapolation gives large differences in the high-frequency region,
even larger than the precision of the different codes while, as
expected, the effect is small for low-order modes.  If the
differential equations are solved with $r/P$ as integration variable,
an almost increased difference is introduced, always lower than $0.001
\,\mu$Hz, i.e., of the order of the precision among most of the codes
using the same numerical techniques. The use of the Lagrangian
perturbation to the pressure as variable introduces the same
oscillatory pattern as was obtained with \LOSC.

%-------------------------------------------------------------------
\section{Computational variations}

%..............................................
\subsection{The influence of the gravitational constant $G$}
\label{sec:grav}

As indicated in Table~\ref{tab:2} different values of $G$ are used by
different codes. 
In many cases the value differs from the value,
$G = 6.67232 \times 10^{-8}$\,cgs, which was used in the computation of
model M4k.
As a result, as seen by the pulsation code the equilibrium model is not
strictly in hydrostatic equilibrium, thus potentially causing errors in
the computed frequencies.

\begin{figure}
\centering
%\begin{tabular}{cc}
  \rotatebox{-90}{\includegraphics[width=6cm]{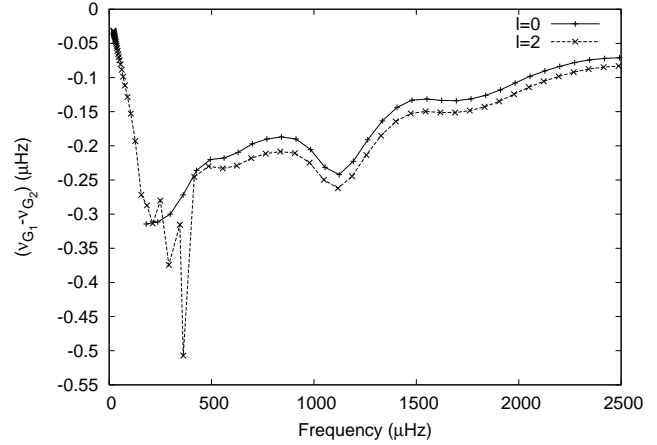}}
%\end{tabular}
\caption{Frequency comparison of modes with $\ell=0$ and 2 obtained
  with \GRACO,($p^\prime$, $\delta P=0$, no Richardson extrapolation, $r$)
  and the
  model M4k when two different values of the gravitational constant $G$ are
  used.}
\label{fig:25}       % Give a unique label
\end{figure}

In this section we examine the consequence of such inconsistencies.
%between the values of the  gravitational constant $G$, used in
%the oscillation code  and in the code computing the equilibrium model.
%Most of the oscillations code use their own value of  the gravitational
%constant $G$, independently of the one used in computing the model.
Thus we test the influence of the choice of a particular
gravitational constant $G$, within always the recent experimental
values found in the literature. The value of this constant is not very
accurately known, and the different values we can find in the
literature can have an impact on the frequency calculation of the same
order or even larger than the differences studied here. To illustrate
this, two extreme values of $G$ found in the literature have been
chosen: $G_1=6.6716823\times 10^{-8}$ cgs (as fixed for Task~1), and
$G_2=6.693\times 10^{-8}$ cgs \citep{g}, the most recent one,
although with a quoted random error of $\pm 0.027 \times 10^{-8}$ and a 
systematic error of $\pm 0.021 \times 10^{-8}$ cgs it is consistent
with the previous value.
The present recommended value of the Committee on Data for Science and
Technology (CODATA) can be found in the World Wide Web at
{\tt physics.nist.gov/constants},
and it is closer to that fixed in Task~1.
The comparison has been carried out with \GRACO\ at fixed ``degrees
of freedom" ($P^\prime$, $\delta P=0$, no Richardson extrapolation, $r$). The
differences obtained using both values of $G$ are shown in
Fig. \ref{fig:25} for modes with $\ell=0$ and 2. The equilibrium model
used is M4k.

Surprisingly, the differences obtained in the region around the
fundamental radial mode are much larger than those obtained for the
direct frequency inter-comparison for the different codes or for the
same code using different integration numerical schemes. As expected
for radial modes the largest difference is obtained for the
fundamental radial mode. The value of this difference
(more than $0.3 \,\mu$Hz in magnitude) is really considerable.
In the asymptotic regions the
influence of the value of this parameter decreases until reaching
values lower than $0.1 \,\mu$Hz. Nevertheless, these differences are
larger than those obtained by using or not the Richardson
extrapolation. The presence of wiggles for
$\ell=2$ for the mixed modes is also remarkable.

The relative difference between $G_1$ and $G_2$ is around $3 \times 10^{-3}$.
In contrast, the relative difference between, say, the value chosen for
Task~1 and the \ADIPLS\ value is only $5 \times 10^{-5}$.
Thus we might expect this difference to cause a frequency difference of order
$5 \times 10^{-3} \, \mu$Hz.
In fact, both the magnitude and shape of the differences between \ADIPLS\ 
and \GRACO\ in the middle panel of Fig.~\ref{fig:4} are consistent with this
estimate.
In general, as already discussed, it appears that much of the differences 
between codes using the same numerical scheme results simply from the
different values of $G$.
Thus, it is obviously important to ensure that consistent values of
$G$ are used in the evolution and oscillation calculations.

%..............................................
\subsection{The choice of dependent variables and equations}
\label{sec:dep}

The most dramatic difference between the different codes is the oscillatory
variation shown by the computations using the LAWE or $\delta P$, relative
to the reference \GRACO\ results (e.g., Fig.~\ref{fig:4}).
The oscillatory nature and the `period' of the variation indicated that
it reflected a sharp difference in some aspect of the model, located
approximately in the region of the second helium ionization zone.
After various failed attempts to identify the cause of the variation it
transpired that it resulted from a, previously known
\citep[e.g.,][]{Boothr2003}, inconsistency in the OPAL equation-of-state
tables \citep[the original tables of][]{Rogers1996}
used in the computation of model M4k with the
\ASTEC\ code.

To understand this we consider the computation of 
\begin{equation}
A^* = {1 \over \Gamma_1} {\dd \ln P \over \dd \ln r} 
- {\dd \ln \rho \over \dd \ln r} \; .
\label{eq:arho}
\end{equation}
  From the point of view of stellar evolution calculation this is somewhat
inconvenient, since the gradient of $\rho$ does not appear directly
in the equations of stellar structure.
Thus evaluation of $A^*$ would involve numerical differentiation of $\rho$.
A potentially more convenient formulation is obtained by rewriting
the expression in terms of the temperature gradient which is known from
the equation of energy transport, given that $\rho$ is a function,
known from the equation of state, of pressure, temperature and composition
which we characterize solely by the hydrogen abundance $X$.
Thus we obtain
\begin{eqnarray}
A^* &=& - \left({ \partial \ln \rho \over \partial \ln T} \right)_{P, X}
(\nabla - \nabla_{\rm ad}) {\dd \ln P \over \dd \ln r} \nonumber \\
&& + \left({ \partial \ln \rho \over \partial \ln X} \right)_{P, T}
{\dd \ln X \over \dd \ln r} \; ;
\label{eq:atemp}
\end{eqnarray}
here, as usual, $\nabla = \dd \ln T / \dd \ln P$ and $\nabla_{\rm ad}$ is
its adiabatic value.
This expression still requires numerical differentiation of $X$,%
\footnote{except if diffusion is taken into account; in that case
the gradient of $X$ appears directly in the equations (see, for example,
the description of \ASTEC\ by \cite{astec})}
but in much of the model $X$ varies little or not at all and hence the term in
$\dd \ln X/\dd \ln r$ makes a limited contribution.
Furthermore, in the bulk of convective zones, with homogeneous composition,
Eq.~\ref{eq:atemp} accurately reflects the small value of $A^*$ resulting
from the small value of $\nabla - \nabla_{\rm ad}$ determined, e.g., from
the mixing-length treatment.

\begin{figure}
\centering
%\begin{tabular}{cc}
  {\includegraphics[width=8cm]{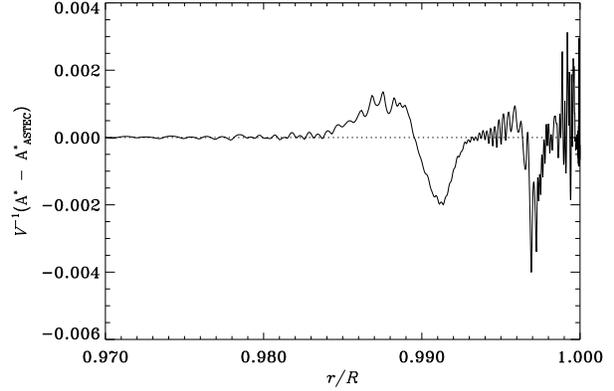}}
%\end{tabular}
\caption{Differences between $A^*$ as evaluated from Eq.~\ref{eq:arho}
and evaluated from Eq.~\ref{eq:atemp} as done in \ASTEC, for model M4k,
against fractional radius.
The differences have been normalized by $V = |\dd \ln P / \dd \ln r|$.
  }
\label{fig:compa}       % Give a unique label
\end{figure}

The transformation in Eq.~\ref{eq:atemp} obviously assumes that the thermodynamical
quantities used are mutually consistent.
As noted by \citet{Boothr2003}, this is not the case of the original OPAL tables
provided by \citet{Rogers1996}.
The effect of this on $A^*$ is illustrated in Fig.~\ref{fig:compa} 
for model M4k, in terms
of the difference between the value of $A^*$ calculated directly from 
Eq.~\ref{eq:arho}, using numerical differentiation to evaluate 
$\dd \ln \rho/ \dd \ln r$,%
\footnote{at this level of precision it does not matter whether
$\dd \ln P / \dd \ln r$ is calculated from the equation of hydrostatic support
or through numerical differentiation.}
and the value, $A^*_{\rm \ASTEC}$, evaluated from Eq.~\ref{eq:atemp}.
There are obvious systematic differences,%
\footnote{the small-scale rapid variation is probably associated
with the interpolation procedure in the equation-of-state tables}
concentrated in the ionization zones of hydrogen and helium.
These would indeed affect the oscillations as a `sharp feature' and hence
cause an oscillatory signature in the frequencies, with the `period'
seen in Fig.~\ref{fig:4} and elsewhere
\citep[e.g.,][]{Gough1990, Montei2005, Houdek2007}.

To test further this interpretation of the results we have recomputed frequencies
for model M4k, but replacing $A^*_{\rm \ASTEC}$ with $A^*$ computed from
Eq.~\ref{eq:arho}.
For this modified model the differences between \ADIPLS\ and \LOSC\ are smaller by
an almost an order of magnitude than the differences illustrated in Fig.~\ref{fig:4}
and with none of the systematic oscillatory character.

We finally note that transforming between the equations using $P'$ and
$\delta P$ also depends on the equation of hydrostatic equilibrium and
hence on a consistent choice of $G$ in the oscillation equations
(cf. Sect.~\ref{sec:grav}).
However, within the actual range of $G$-values used by the different codes
this effect is smaller than the effect of the inconsistency in $A^*$.

%..............................................
\subsection{The choice of independent variable}
\label{sec:indep}

\begin{figure}
\centering
  \rotatebox{-90}{\includegraphics[width=6cm]{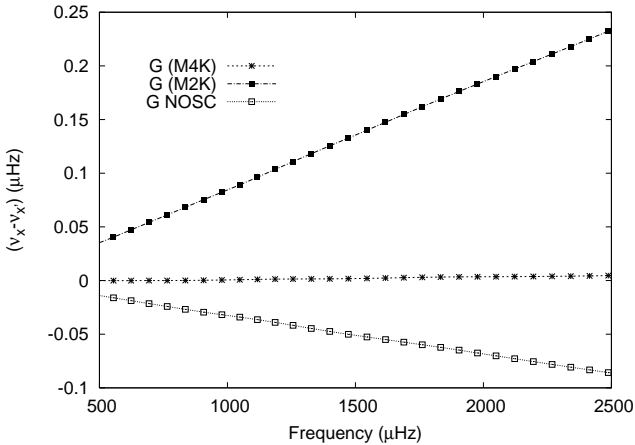}}
\caption{Differences between the use of $r$ or $r/P$ as integration
  variable as a function of the mode frequency, 
  for radial modes.
%  \note [I assume!!]
  Different $G$ values
  are used in the \GRACO\ oscillation code: $G$ (M4K) is the $G$ value
  used to obtain the M4K model, $G$ (M2K) is the $G$ value used to
  obtain the M2K model, and $G$ \NOSC\ is the $G$ value used by the
  \NOSC\ code. M4K has been used as equilibrium model for this comparison.}
\label{fig:comprp}       % Give a unique label
\end{figure}

We have found a fairly significant dependence on the $G$ value used by
the oscillation codes for the differences of using the integration
variables $r$ or $r/P$. Fig. \ref{fig:comprp} shows this dependence
in the case of $\ell = 0$.
%\note [I assume!!!]
In this figure, the differences between the use of $r$ or $r/P$ with the
three different $G$ values presented in Table 2, are depicted. All
of them have been obtained using \GRACO\ and M4K.
The differences increase with frequency, and their value
depend on the difference $G_{eq}-G_{osc}$, where $G_{x}$ is the $G$
value used to obtain the equilibrium model or the oscillation
frequencies, respectively.
In fact, the transformation between using $r$ and $r/P$ as independent
variables uses the equation of hydrostatic equilibrium and hence the
result is obviously sensitive to whether or not a consistent value of $G$
is used.
By far the smallest differences are obtained when
the same $G$ are used in the equilibrium and oscillation codes.
%\note [I do not understand the meaning of the following sentence; 
%perhaps it is best dropped].
%This
%behaviour is not present when the rest of the possible numerical
%resolutions are used.

Even when the same value of $G$ is used, we have found a slight 
effect of using $r/P$ as independent variable, as
implemented in a few of the codes (see also \citet{nosc}), instead of $r$;
in Fig.~\ref{fig:comprp} the maximum difference is $0.008\, \mu$Hz,
as also found in Fig.~\ref{fig:4}.
The origin of 
choosing $r/P$ seems to be a concern that near the surface, where $r$ varies
little between adjacent mesh points, rounding errors in the evaluation
of differences in $r$, when representing the differential equations on
finite-difference form, could have a significant effect on the
results.  Given the rapid variation of $P$ in the near-surface layers
this problem is obviously avoided if $r/P$ is used instead.  While
there might well be such problems with using $r$ when variables are
represented in single-precision form (with four-byte reals) it seems
unlikely to be a problem when using double-precision variables.%
\footnote{In contrast, the problem would be severe if the interior
mass were used as independent variable.}  On the other hand, the
present test has been carried out with models transferred in the {\tt
FGONG} format, where variables are given with ten significant digits.
Here one cannot exclude that rounding errors might be significant.

To test this JC-D computed frequencies with \ADIPLS\ using both the
original binary (double-precision) version of M4k and the version
resulting from conversion to {\tt FGONG} format.  The frequency
differences between these two cases were always less than $10^{-5} \,
\mu$Hz, strongly indicating that the use of the {\tt FGONG} format is
not a concern with the present level of accuracy.

%\note [Following our discussion I have made some cuts in this paragraph;
%in particular, without a response from Janine I am not convinced that the
%reference to her paper is relevant here.
%I may change my mind once we hear from her, and possibly in time for 
%a revision following refereeing!].
It is obvious that the choice of independent variable also affects the
{\it truncation\/} error, i.e., the error in the representation of the
differential equations on finite-difference form.
%\citep{nosc}.
%together with the use of linear oscillation equations, 
This can presumably be the reason for the differences found with \GRACO\ when
comparing the use of $r$ and $r/P$ (e.g., Fig.~\ref{fig:18}),
although the effects are small.
%and different $G$ values are used.
Further analysis is required
to decide which is the more accurate representation.

%-------------------------------------------------------------------
\section{Conclusions.}

A complete inter-comparison of the frequencies calculated by using a
set of oscillation codes has been presented.  Nine oscillation codes
have been used: \ADIPLS, \FILOU, \NOSC, \LOSC, \GRACO, \OSCROX, \PULSE, \POSC\ and \ROMOSC. The only free parameter for each code has been the
integration numerical scheme. In order to ensure that the same physics
has been imposed to all the codes two equilibrium models of a
$1.5M_\odot$ star have been supplied, with 4042 (M4k) and 2172 (M2k)
mesh points in their grids, respectively. No re-meshing has been
allowed.  The model M2k pre\-sents inadequate numerical resolution in
the Brunt-V\"ais\"al\"a frequency close to the boundary of the
convective core.  The model M4k does not suffer from these problems.
In the present paper we have used these models only to show the
sensitivities of the oscillation codes to possible numerical
inaccuracies in the equilibrium models.  A complete study of the
accuracies of the different evolutionary codes is provided by Lebreton
et al. in this volume.

Several inter-comparisons have been performed: 1) Direct frequency
inter-comparison, 2) large separation (LS), 3) small separation (SS),
and 4) different experimental values of the gravitational constant
$G$.

The main conclusions can be summarized as:

\begin{itemize}

\item When codes using the same numerical integration sche\-mes are
  compared, the general precision obtained for model M4k
  ($0.02 \,\mu$Hz) is higher than the expected precision of the observational
  data obtained with the CoRoT mission.
  This precision is always one order of magnitude better or even
  more when compared with that obtained for model M2k
  ($0.5 \, \mu$Hz). The largest differences are obtained for the 
  high-frequency region.

\item The use of a second-order integration scheme plus Richardson
   extrapolation or a fourth-order integration scheme does not improve
   the agreement between the codes when compared with the use of a second-order
   scheme. In the cases here presented it was not possible to
   distinguish between the use of a fourth-order scheme
   or Richardson extrapolation,
   which always give very acceptable consistency for model M4k.

\item However, the use of a fourth-order integration scheme (or a second-order
  plus Richardson extrapolation) introduces
  differences larger than $0.1 \,\mu$Hz ($1.5 \,\mu$Hz for M2k and 
  $0.8 \,\mu$Hz for M4k) compared with a second-order scheme,
  mainly located in the regions of high order (high frequency for p-modes
  and low frequency for g-modes).
  This indicates that the second-order schemes have inadequate numerical
  precision on the mesh provided in M4k and, obviously, even more so on
  the sparser mesh in M2k.

\item For ``large separations" the agreement between different
  comparable codes using model M4k,
  with every possible numerical scheme is generally better than the
  expected precision of the data.
  However, the equation-of-state inconsistency in model M4k 
  (see Sect.~\ref{sec:dep})
  introduces a signature that it potentially significant.
  The use of the Richardson
  extrapolation together with a second-order integration scheme, or a
  fourth-order scheme, gives differences
  for high-order p- and g-modes
  relative to the use of only a second-order scheme
  that are comparable to the expected observational accuracies, 
  indicating that the accuracy of the latter scheme is inadequate.

\item The ``small separations'' present differences for model M4k
  lower than the expected observational accuracies everywhere,
  although the inconsistency in M4k has a noticeable effect.
% Only the use of LAWE to obtain the radial modes can gives differences for
% $\delta_{02}$ close to these accuracies.
  Use of Richardson extrapolation has a potentially significant effect,
  indicating that the precision of second-order schemes may be inadequate.

\item The use of $r/P$ as integration variable instead of $r$ has a
  small influence when a consistent value of $G$ is used.
  However, an inconsistency between the $G$
  values used to obtain the equilibrium model and the oscillation
  frequencies can give substantial differences between the use of these
  integration variables, over the range of $G$ values in Table 2.

\item For model M2k, the mixed modes near the avoided crossing
  present large differences, reaching even up to $4 \,\mu$Hz. These
  differences are not present when model M4k is used.
% The only reason can not be the increment of mesh points in the inner zones. 
  The problems with $N^2$ in M2k 
  are undoubtedly the main source of this difference.
  The same behaviour is found for the trapped g-modes. Therefore, the
  correct numerical description of $N^2$ is critical for the value of
  the frequencies of these trapped modes or in avoided crossing.

\item The use of $\delta P$ or $P^\prime$ as
  eigenfunction, or the solution of the LAWE, may have a significant effect
  if the equilibrium model is based on thermodynamic quantities that are not
  internally consistent, as is the case for the OPAL tables used to calculate
  model M4k.
  Further effects arise
  when there are problems with the Brunt-V\"ais\"al\"a frequency
  (as for model M2k); here a
  different choice can change the frequencies of some modes since
  $N^2$ does not appear in the differential equations when $\delta P$
  or the LAWE are used and it does when using $P^\prime$.

\item The value of the gravitational constant $G$ 
  in the oscillation calculations can introduce
  non-negligible differences as well, if it is not the same as the value
  used in the equilibrium model. When the two extreme values found in
  the literature are used,
  such inconsistency yields
  differences in the range $[-0.35,-0.08] \,\mu$Hz 
  for model M4k, larger than those obtained when
  different numerical integration schemes are used.
  Differences between the values of $G$ actually used by the different codes,
  although less extreme,
  account for much of the difference between the computed frequencies.

\end{itemize}

Therefore, for a proper pulsational study, we require that 
the number of mesh points and their distribution must be such as to yield
an equilibrium model that satisfies the dynamical equations with
sufficient accuracy also in the regions of the star where the 
physical quantities present rapid variations
(e.g., the outer layers and $\mu$-gradient zones).
In addition, the mesh used in the pulsation calculation must properly
resolve the eigenfunctions of the highest-order modes considered.
In the present case model M4k, with 4042 points, appears to satisfy these
conditions although the use of a fourth-order integration scheme, or
a second-order scheme and Richardson extrapolation, is still needed in the
oscillation calculation;
these higher-order schemes give significant improvements compared with
the use of a simple second-order scheme.
The use of a second-order integration plus Richardson
extrapolation scheme is not distinguishable, in accuracy and
precision, from the use of a fourth-order integration scheme.
A correct physical and numerical description of the
Brunt-V\"ais\"al\"a frequency is essential when $P^\prime$ is used as
eigenfunction;
in particular, inconsistencies in the equation of state can have 
serious effects on the frequencies.
Inconsistency between the value of $G$ used in the oscillation calculation
and the value used to compute the equilibrium model,
within the range of the different values of $G$ found in the literature,
may lead to substantial errors in the computed frequencies.

We note that the situation is somewhat different if consistent values of $G$
are used in the evolution and oscillation calculations.
Then the effect on the frequencies is approximately given, according to
homology arguments, as a scaling by $(GM)^{1/2}$.
However, since the product $G M_\odot$ is known extremely precisely
from planetary motion in the solar system, any variation in $G$ should be
reflected in a corresponding change in the assumed value of $M_\odot$.
If this is the case, and if the model is characterized by a given value
of $M/M_\odot$ (as is typically the case) the effect on the frequencies
of changes in $G$ are very small \citep[see also][]{Christ2005}.

In further tests more care is required to secure the full consistency 
of the models: a consistent equation of state should be used, and the
value of $G$ should obviously be the same in the equilibrium-model and
the pulsation calculations; indeed, this strongly argues for including the value
of $G$ as one of the parameters in the model file.
The main conclusion of this extensive investigation, however, it positive:
with a properly resolved equilibrium model the broad range of oscillation
codes likely to be involved in the asteroseismic analysis of data from
CoRoT and other major upcoming projects generally give consistent results,
well within the expected errors of the observations.
Thus, although the remaining problems in the calculation evidently require
attention, we can be reasonably confident in our ability to compute
frequencies of given models and hence in the inferences 
concerning stellar structure drawn from comparing
the computed frequencies with the observations.

%-------------------------------------------------------------------
\begin{acknowledgements}
This work was supported by the Spanish PNE under Project number ESP
2004-03855-C03-C01, and by the European Helio- and Asteroseismology
Network (HELAS), a major international collaboration funded by the
European Commission's Sixth Framework Programme. A.M. and
J.M. acknowledge financial support from the Belgian Science Policy
Office (BELSPO) in the frame of the ESA PRODEX 8 program (contract
C90199). MJPFGM is supported in part by FCT and FEDER (POCI2010)
through projects {\small POCI/CTE-AST/57610/2004} and {\small
POCI/V.5/B0094/2005}.
\end{acknowledgements}

%-------------------------------------------------------------------
% BibTeX users please use
%\bibliographystyle{spmpsci}

%*******************************************************************
\end{document}